\documentclass[12pt]{article}

\usepackage{slashed}
\usepackage{color, verbatim}
\usepackage{latexsym}
\usepackage{amsmath}
\usepackage{graphicx}
\usepackage{arydshln}
\usepackage[dvipsnames]{xcolor}
\usepackage{adjustbox}
\usepackage{bbm}

\usepackage{cite}
\usepackage{hyperref}

\usepackage{amssymb}
\usepackage{graphicx}
\usepackage{bm}
\usepackage{hyperref}
\usepackage{adjustbox}
\usepackage{booktabs}
\usepackage{multirow}
\usepackage{rotating}
\usepackage{placeins}
\makeatletter

\@addtoreset{equation}{section}
\makeatother

\newcommand{\mc}[1]{\textcolor{blue}{#1}}

\newcommand{\ctoprule}{\toprule[0.5mm]}
\newcommand{\cbottomrule}{\bottomrule[0.5mm]}
\newcommand{\cmrule}{\midrule[0.25mm]}

\newcommand{\vtext}[1]{\begin{sideways}\small{#1}\end{sideways}}

\newcommand{\ifb}{\text{ fb}^{-1}}
\newcommand{\ipb}{\text{ pb}^{-1}}
\newcommand{\mev}{\,\text{MeV}}
\newcommand{\gev}{\,\text{GeV}}
\newcommand{\tev}{\,\text{TeV}}
\newcommand{\ope}{\mathcal{O}}

\newcommand{\etmiss}{E_T^\text{miss}}

\DeclareMathOperator{\arccoth}{arcCoth}

\graphicspath{{figures/}}

\makeatletter
\g@addto@macro\bfseries{\boldmath}
\makeatother

\allowdisplaybreaks

\begin{document}
\thispagestyle{empty}
\begin{flushright}
IPPP/20/25 \\
\end{flushright}
\vspace{0.8cm}

\begin{center}

\begin{center}

%\vspace{.5cm}

{\Large\sc The effective field theory of low \\[0.2cm] 	scale see-saw at 
colliders}

\end{center}

\vspace{0.8cm}

\textbf{
Anke Biek\"otter$^{\,a}$, Mikael Chala$^{\,b}$ and Michael Spannowsky$^{\,a}$}\\

\vspace{1.cm}
{\em {$^a$Institute for Particle Physics Phenomenology, Department of Physics, Durham University, South Road, 
Durham DH1 3LE, United Kingdom}}\\[0.2cm]
{\em {$^b$CAFPE and Departamento de F\'isica Te\'orica y del Cosmos,
Universidad de Granada, E–18071 Granada, Spain}}\\[0.2cm]
\vspace{0.5cm}

%\end{comment}

\end{center}

\begin{abstract}
We study the Standard Model effective field  theory ($\nu$SMEFT)
extended with operators involving right-handed neutrinos, 
focussing on 
the regime where the right-handed neutrinos decay promptly on 
collider scales to a photon and a Standard Model neutrino. 
This scenario arises naturally for right-handed neutrinos with 
masses of the order $m_N \sim 0.1 \dots 10\, \gev$. We limit the relevant 
dimension-six operator coefficients using 
LEP and LHC searches with photons and missing energy in the final 
state as well as pion and tau decays. 
While bounds on new physics contributions are generally in the TeV scale for 
order one operator coefficients, 
some coefficients, however, remain very poorly constrained or even entirely
evade bounds from current data. 
Consequently, we identify such weakly constrained scenarios and propose new searches for rare top and tau decays 
involving photons to 
probe potential new physics in the $\nu$SMEFT parameter space. 
Our analysis highlights the importance of performing dedicated searches for new rare tau and top decays.
\end{abstract}

\newpage

\tableofcontents

\section{Introduction}
The discovery that neutrinos are 
massive~\cite{Fukuda:1998mi,Toshito:2001dk,Giacomelli:2001td,Fukuda:2001nj,
Fukuda:2001nk, Ahmad:2002jz, Ahmad:2002ka} is direct evidence of physics beyond 
the Standard 
Model~(SM). One of the most appealing explanations of neutrino masses is the so 
called type I 
\textit{see-saw} mechanism~\cite{Minkowski:1977sc,Yanagida:1979as,GellMann:1980vs,
Mohapatra:1979ia}. In its most simple incarnation, it 
extends the SM with three right-handed (RH) neutrinos~$N$ with very large 
lepton-number violating (LNV) Majorana mass terms $\sim m_N \overline{N^c}N$ and 
$\mathcal{O}(1)$~Yukawa couplings~$y$ for the SM neutrinos~$\nu$; 
so that the mass of the latter is $m_\nu\sim y 
v^2/m_N\lesssim$ eV with $v\sim 246$ GeV being the Higgs vacuum expectation 
value (VEV). 

There are however two big modifications of this simplistic setup, both of which 
can hold simultaneously: $(i)$ \textit{a priori}, at least one field $N$ can be 
at the electroweak (EW) scale\footnote{This requires $y\ll 1$, which is still natural in the t'Hooft 
sense~\cite{tHooft:1979rat}.}~\cite{Pilaftsis:1991ug,Borzumati:2000mc}. 
$(ii)$ $N$ can be part of a bigger sector with further heavy particles 
not related to LNV; this is in general the case in left-right symmetric 
inspired models~\cite{Dev:2016dja}, GUTs~\cite{Dev:2009aw} and 
others~\cite{Borzumati:1986qx}. If both $(i)$ and 
$(ii)$ hold simultaneously, the physics at the EW scale can be described by an 
effective field theory (EFT) involving not only the SM degrees of freedom but 
also light sterile neutrinos. This EFT is known as $\nu$SMEFT.

$\nu$SMEFT was first considered in 
Refs.~\cite{delAguila:2008ir,Aparici:2009fh}; see also 
Ref.~\cite{Bhattacharya:2015vja}. The 
first complete and non-redundant basis of operators of up to 
dimension six was provided in Ref.~\cite{Liao:2016qyd}. $\nu$SMEFT-operators relevant 
at energies 
below the EW scale, where the top and the $W$, $Z$ and Higgs bosons are 
integrated out, has been recently studied, including partial renormalization 
group equations, in Ref.~\cite{Chala:2020vqp}. The corresponding chiral EFT 
valid at 
energies below the QCD scale has been considered in Ref.~\cite{Dekens:2020ttz} 
for operators 
relevant for neutrinoless double beta decay. 
%See also Ref.~\cite{} for a recent 
%work on the flavour structure of the $\nu$SMEFT. 

The collider phenomenology of $\nu$SMEFT has been explored in a variety of 
studies, which can be classified depending on the interactions which are assumed to 
trigger the production and decay of $N$. Most works have focused on the decay of 
$N$ via active-sterile neutrino mixing, but in 
general 
both production and decay can be mediated and even dominated by effective operators~\cite{Cai:2017mow}. The 
parameter space in which they both occur via tree-level generated
contact interactions has been studied in Refs.~\cite{delAguila:2008ir,Duarte:2015iba}. This regime arises 
naturally when there are no electrically charged particles in the UV. The 
parameter space in which effective bosonic 
operators dominate both production and decay has been considered in 
Ref.~\cite{Butterworth:2019iff}. This regime is inherent to models in which the 
new physics 
undergoes a $\mathbb{Z}_2$ symmetry that forbids tree-level operators in the 
EFT; see Ref.~\cite{Chala:2020vqp} for an example including a thorough 
calculation of all 
operators arising in the one-loop matching.

However, the most general scenario is that in which both tree-level 
as well as loop-induced operators 
are generated when integrating out the new physics that manifests itself as particles in the UV. In that case, four-fermion 
operators trigger the production of $N$ at colliders, which subsequently decays 
to $N\to\nu\gamma$ via bosonic operators. No systematic study of $\nu$SMEFT 
in this more likely regime has been performed, beyond some preliminary exploration
of potential displaced vertex signals~\cite{Duarte:2015iba,Duarte:2016caz}.
We intend to fill this gap in this article.

In section~\ref{sec:parameter_space} we define and discuss the parameter space of $\nu$SMEFT that we are interested in. 
In section~\ref{sec:bosonic}, we present limits on the bosonic operators 
of our $\nu$SMEFT Lagrangian. 
In section~\ref{sec:searches}, we study the impact of 
four-fermion operators in 
$\nu$SMEFT on different experimental signatures, focussing on 
LHC searches for one lepton, one photon and missing 
energy at CMS in section~\ref{sec:lepton_photon_etmiss}. 
In section~\ref{sec:two_photons_etmiss}, 
we discuss searches for two photons and missing energy at the same experiment. 
In sections~\ref{sec:pion} and \ref{sec:tau} we explore the implications of the 
EFT on pion and tau decays, respectively.  
Finally, in section~\ref{sec:LEP}, we investigate the $\nu$SMEFT contributions 
to processes with one or multiple photons and missing energy in the final state
as studied by the LEP experiments. 

On the basis of these 
results, 
in section~\ref{sec:limits} we summarise the obtained limits on the $\nu$SMEFT operators, thereby 
unravelling which directions in parameter space are less or not yet constrained and therefore 
identifying where new physics is more likely to be found. In section~\ref{sec:top}, we develop new 
search strategies to explore these not yet constrained directions, which 
include operators triggering $\tau$~decays via the $\tau \to \ell \gamma \gamma \nu (\nu)$ channel
 and the rare top decay $t\to \gamma 
b\ell \nu$. 
We conclude in section~\ref{sec:conclusions}.

\section{Relevant parameter space}
\label{sec:parameter_space}

%
%%%%%%%%%%%%%%%%%%%%%%%%%%%%%%%%%%%%%%
%Basis of dim 6
\begin{table}[t]
\begin{center}
\begin{tabular}{c c l c c l }
\ctoprule
%&Operator& Notation& &Operator&Notation\\
%\cmidrule{2-3}\cmidrule{5-6}
% Dimension 6
% 4F: LLLL
%
\multirow{4}{*}{\vtext{SF}}&\multicolumn{5}{c}{$(\overline{L}N)\tilde{H} 
(H^\dagger H)\,\,\,\,$ ${\cal O}_{LNH}$ (+h.c.)}\\
&$(\overline{N}\gamma^\mu N)(H^\dagger i \overleftrightarrow{D_\mu} H)$&${\cal 
O}_{HN}$&&
$(\overline{N}\gamma^\mu e)(\tilde{H}^\dagger i D_\mu H)$&
${\cal O}_{HNe}$ 
(+h.c.)\\
&$(\overline{L}\sigma_{\mu\nu} N)\tilde{H}B^{\mu\nu}$&${\cal O}_{NB}$ 
(+h.c.)&&$(\overline{L}\sigma_{\mu\nu} N)\sigma_I\tilde{H}W^{I \mu\nu}$&${\cal 
O}_{NW}$ (+h.c.)\\[0.1cm]
\cmrule
\multirow{3}{*}{\vtext{RRRR}}&$(\overline{N}\gamma_\mu N)(\overline{N}\gamma^\mu 
N)$&${\cal O}_{NN}$& &&\\
&$(\overline{e}\gamma_\mu e)(\overline{N}\gamma^\mu 
N)$&${\cal O}_{eN}$&&$(\overline{u}\gamma_\mu 
u)(\overline{N}\gamma^\mu N)$&${\cal O}_{uN}$\\
&$(\overline{d}\gamma_\mu d)(\overline{N}\gamma^\mu N)$&${\cal 
O}_{dN}$&&$(\overline{d}\gamma_\mu u)(\overline{N}\gamma^\mu 
e)$&${\cal O}_{duNe}$ (+h.c.)\\[0.1cm]
\cmrule
% 4F: RRRR
\multirow{1}{*}{LLRR}&$(\overline{L}\gamma_\mu L)(\overline{N}\gamma^\mu 
N)$&${\cal O}_{LN}$&&$(\overline{Q}\gamma_\mu Q)(\overline{N}\gamma^\mu 
N)$&${\cal O}_{QN}$\\[0.1cm]
\cmrule
% 4F: LLRR and LRRL
\multirow{2}{*}{\vtext{LRRL}}&$(\overline{L} N)\epsilon 
(\overline{L}e)$&${\cal O}_{LNLe}$ (+h.c.)&&$(\overline{L} 
N)\epsilon (\overline{Q} d)$&${\cal O}_{LNQd}$ (+h.c.)\\
&$(\overline{L}d)\epsilon (\overline{Q} N)$&${\cal O}_{LdQN}$ 
(+h.c.)&&$(\overline{Q}u)(\overline{N}L)$&${\cal O}_{QuNL}$ (+h.c.)\\[0.1cm]
\cbottomrule
\end{tabular}
\caption{Lepton number conserving operators 
containing a 
RH neutrino $N$~\cite{Liao:2016qyd}.}\label{tab:basis}
\end{center}
\end{table}
%%%%%%%%%%%%%%%%%%%%%%%%%%%%%%%%%%%%%%
%
%
The renormalizable Lagrangian of $\nu$SMEFT reads
\begin{align}
 L_{SM+N} = K - V -\bigg\lbrace\overline{Q} \mathbf{Y_d} H d 
&+\overline{Q}\mathbf{Y_u}\tilde{H}u +\overline{L} \mathbf{Y_e} 
H e 
+ \overline{L} \mathbf{Y_N}  \tilde{H} N + 
\frac{1}{2} \overline{N^c} \mathbf{M_N}N\bigg\rbrace\,,
\end{align}
where $K$ and $V$ are the kinetic terms and the scalar potential, 
respectively, while $L$ and $Q$ represent the left-handed (LH) lepton and quark doublets, 
respectively. Accordingly, $e$ and $u$ and $d$ stand for the  
 RH leptons and the up and down quarks. We use the 
symbol $H$ to 
 denote the Higgs doublet, while $\tilde{H} = \epsilon H^*$ with 
 $\epsilon$ being the 
 fully antisymmetric tensor in two dimensions. $B_{\mu\nu}$ and 
 $W^I_{\mu\nu}$ represent the weak field strength tensors. Flavour indices 
 are not shown explicitly. Without loss of generality, we work on the basis
 in which the Yukawa matrices $\mathbf{Y_e}$ and $\mathbf{Y_d}$ are diagonal.

The effective Lagrangian 
can be parameterised as
\begin{equation}
 L_{EFT} =   
\frac{1}{\Lambda}(\alpha_{NNH} \, \mathcal{O}_{NNH} + \alpha_{NNB}\, \mathcal{O}_{NNB})+\frac{1}{\Lambda^2}
\sum_i\alpha_i \, \mathcal{O}^6_i\,,
\end{equation}
where 
$\mathcal{O}_{NNH} = \overline{N^c} N H^\dagger H$ and $\mathcal{O}_{NNB} =\overline{N^c}\sigma_{\mu\nu} 
N B^{\mu\nu}$ are the (LNV) dimension-five operators, and 
$\mathcal{O}^6$ represent the dimension-six operators in Tab.~\ref{tab:basis}.

Following the line of thought of Ref.~\cite{Cirigliano:2005ck}, we assume that 
the RH 
neutrino mass term 
$\mathbf{M_N}$ is the only source of LNV. Likewise, in order to reduce the plethora of independent Wilson coefficients in the EFT,
we assume universality in $N$ and no off-diagonal couplings. We also assume 
universality and no flavour violation in the light lepton sector.
%{\bf the following sentence was weird and maybe still is...} Finally, we 
%require no flavour violating operators in relation to the quarks with flavour 
%universality in the light sector. 
Finally, we require flavour universality in the light quark sector and
no flavour-violating transitions between any of the three quark families.
Most of these assumptions, if not all,  can be 
enforced by symmetries; \textit{e.g.} $e\leftrightarrow\mu$ for light lepton 
flavour universality. We refer the reader to the Appendix~\ref{app:lag} for the 
explicit expressions of the Lagrangian, including all flavour indices.  

As an example, let us show the full structure of the operators $\ope_{eN}$ 
and $\ope_{uN}$
\begin{equation}
\begin{split}
 \alpha_{eN} \, \mathcal{O}_{eN} =& \, \alpha_{eN}^{\ell\ell}\, (\mathcal{O}_{eN}^{11ii}+\mathcal{O}_{eN}^{22ii}) 
+\alpha_{eN}^{\ell\tau} \, (\mathcal{O}_{eN}^{13ii}+\mathcal{O}_{eN}^{23ii}+\mathcal{O}_{eN}^{31ii}+\mathcal{O}_{eN}^{32ii})
+ \alpha_{eN}^{\tau\tau} \, \mathcal{O}_{eN}^{33ii}\,, \\
 \alpha_{uN} \, \mathcal{O}_{uN} =&  \,
 \alpha_{uN}^{qq}\, (\mathcal{O}_{uN}^{11ii}+\mathcal{O}_{uN}^{22ii})
 +  \alpha_{uN}^{tt} \, \mathcal{O}_{uN}^{33ii} \, ,
 \end{split}
\end{equation}
where the index $i = \{1, \,  2 ,\, 3\}$ specifies the RH neutrino flavour.

As a consequence of the above conditions, $\mathbf{M_N}$ must be 
proportional to the identity matrix, $\mathbf{M_N} = m_N\mathbbm{1}$, and
$\alpha_{NNH}$ 
and $\alpha_{NNB}$ must vanish~\footnote{Note that if we relax the 
assumption on $\mathbf{M_N}$ being the only source of LNV, $\alpha_{NNH}$ 
would be also proportional to the identity matrix. As such, its effect on the 
RH neutrino mass term could be reabsorbed in the redefinition 
$(\mathbf{M_N})_{ij}\to (\mathbf{M_N})_{ij}+v^2/\Lambda(\alpha_{NNH}^{ij})$. 
The sole effect of $\alpha_{NNH}$ would then appear in the interaction 
$h\overline{N^c}N$. We refer to Ref.~\cite{Butterworth:2019iff} for tests of 
this vertex in Higgs decays.} (likewise for the Weinberg operator). 
The strong constraints from SM neutrino 
dipole moments~\cite{Canas:2015yoa,Miranda:2019wdy},  that would arise upon 
active-sterile neutrino mixing if 
$\alpha_{NB}$ was not vanishing, are therefore evaded.

Without loss of generality, we can make the redefinition
\begin{equation}
(\mathbf{Y_N})_{ij}\to(\mathbf{Y_N})_{ij} + \frac{v^2}{2\Lambda^2}\alpha_{LNH}^{ij} \,.
\end{equation}
Therefore, the effects of the operator $\mathcal{O}_{LNH}$ manifest 
themselves only in rare decays of the Higgs boson; see 
Ref.~\cite{Butterworth:2019iff}. The neutrino mass  matrix then reads
\begin{equation}
 \mathbf{M} = \bigg( \begin{array}{cc}
                      0 &\frac{v}{\sqrt{2}}\mathbf{Y_N}\\
                      \frac{v}{\sqrt{2}}\mathbf{Y_N}^T & \mathbf{M_N}
                     \end{array}\bigg)\,.
\end{equation}
Upon diagonalization, for the active-sterile neutrino mixing matrix one finds
\begin{equation}
 \mathbf{\Theta} = \frac{v}{\sqrt{2}} \mathbf{M_N}^{-1}\mathbf{Y_N}\,.
\end{equation}
 Using the 
Casas-Ibarra parameterization~\cite{Casas:2001sr}, $\mathbf{Y_N}$ can be expressed (remember that we work  in the 
basis in which the charged lepton Yukawa is 
diagonal), as  
\begin{equation}
\mathbf{Y_N} =\frac{\sqrt{2m_N}}{v}\, 
\mathbf{U_{\text{PMNS}}}\, 
\sqrt{\text{diag}}(m_{\nu_1},m_{\nu_2},m_{\nu_3}) \,\mathbf{R}\, 
\approx \frac{\sqrt{m_\nu m_N}}{v}\mathbf{U_{PMNS}}
%\,\bigg(\begin{array}{ccc}
%   0 & 0 & 0 \\
 %  1 & 1 & 1\\
  % 1 & 1 & 1 \\
 % \end{array}\bigg)
,\label{eq:ibarra}
\end{equation}
where 
$\mathbf{U_{PMNS}}$ is the Pontecorvo-Maki-Nakagawa-Nasaka matrix, $m_{\nu_i}$ 
is the mass of the $i$-th SM neutrino and $\mathbf{R}$ is an orthogonal matrix.
In the last step of Eq.~\ref{eq:ibarra}
we have conservatively assumed that 
$\mathbf{R}_{ij}\sim\mathcal{O}(1)$ as well as
$m_{\nu_1}\approx 0$ and $m_{\nu_2}\approx 
m_{\nu_3}=m_\nu$. 

In the following we will show that our model naturally leads to a RH 
neutrino mass-range in which the neutrino decays almost exclusively 
via the $N\to \nu \gamma$ channel and the decay is prompt on collider 
scales.
Under the conservative approximation that 
$\mathbf{U_{PMNS}}_{ij}\sim\mathcal{O}(1)$, the decay
$N\to \ell \ell \nu$ is driven by an off-shell $Z$ boson due to the 
active-sterile 
neutrino mixing. It is approximately given by
\begin{equation}
 \Gamma_\text{mix}(N\to \ell \ell \nu)\approx 
\frac{1}{64\pi^3}\frac{m_\nu}{m_N}\left(\frac{m_N}{v}\right)^4 m_N\,,
\end{equation}
valid for arbitrarily small $m_N$.

On the other hand, tree-level generated 
contact interactions, \textit{e.g.}
$\mathcal{O}_{LNQd}$, drive the decays $N\to qq\nu$, $N\to q q' \ell$ as well as 
purely leptonic decays. The dominant of these modes for the mass range $m_N\gtrsim 1$ GeV is
\begin{equation}
 \Gamma_\text{tree}(N\to q q' \ell) \approx
\frac{N_c\,\alpha_{LNQd}^2}{3072\pi^3}\left(\frac{m_N}{\Lambda}\right)^4 m_N\,,
\end{equation}
and likewise for other four-fermion operators. 

Finally, the decay 
$N\to\nu\gamma$ triggered by the loop-suppressed
$\mathcal{O}_{NA}=c_W\, \mathcal{O}_{NB}+s_W \, \mathcal{O}_{NW}$ is 
\begin{equation}
 \Gamma_\text{loop}(N\to\nu\gamma) =
\frac{\alpha_{NA}^2}{4\pi} \frac{m_N^2 v^2}{\Lambda^4}m_N \,.
\label{eq:loop_decay}
\end{equation}
For $m_\nu\sim 1$ eV, even for $\Lambda\sim 10$ TeV, if 
$\alpha_{NA}\sim 1/(4\pi)$ due to the loop-suppression and $\alpha_{LNQd}\sim 1$, we obtain respectively 
\begin{equation}
\frac{\Gamma_\text{mix}}{\text{GeV}}\approx 10^{-21} 
\left(\frac{m_N}{\text{GeV}}\right)^4 
\,, \quad
\frac{\Gamma_\text{tree}}{\text{GeV}}\approx 10^{-20} 
\left(\frac{m_N}{\text{GeV}}\right)^5 
\,,\quad 
\frac{\Gamma_\text{loop}}{\text{GeV}}\approx 
10^{-15}\left(\frac{m_N}{\text{GeV}}\right)^3
\, .
\end{equation}
Due to the suppression with the LH neutrino mass $\Gamma_\text{mix}$ 
will remain subdominant throughout the whole RH neutrino mass range. 
We therefore obtain the hierarchy $\Gamma_\text{mix} \ll 
\,\Gamma_\text{tree}<\Gamma_\text{loop}$ 
provided $m_N\lesssim 10$ GeV, where for the tree-level decay we take into
account contributions from multiple four-fermion operators.  
%\textit{i.e.} we obtain the hierarchy $\Gamma_\text{mix} \ll 
%\,\Gamma_\text{tree}<\Gamma_\text{loop}$ 
%provided $m_N\lesssim 10$ GeV. 
\footnote{We note that this hierarchy is very different if the 
flavour group is instead $SU(3)^6$ and Minimal Flavour Violation is 
enforced~\cite{Barducci:2020ncz}. The reason is that $\mathcal{O}_{NA}$ 
is no longer 
invariant unless it carries one power of the spurion $\mathbf{Y_N}$, what 
makes 
$\Gamma_\text{loop}$ further suppressed by $\sim m_\nu m_N/v^2$.}

Moreover, to assure that $N$ decays promptly at 
colliders like the LHC,
we require the decay length of $N$ to be
$c\tau \lesssim 4 \,\text{cm}$.
Using Eq.~\eqref{eq:loop_decay}, we find that this is the case for
$m_N\gtrsim 0.04$ GeV and 
$\alpha_{NA} /\Lambda^2 \lesssim 4\pi/(10 \tev)^{2}$.

While the assumption that the RH neutrino decays almost exclusively via the 
$N \to \nu \gamma$ channel sets an upper bound on the RH neutrino mass, 
the requirement of a prompt decay bounds $m_N$ from below, leaving us with
the regime $m_N \in [0.04 , \, 10]\,\gev$ and $\alpha_{NA}\sim\mathcal{O}(1)$ 
for $\Lambda=10$ TeV.
We focus on 
this region of the parameter space hereafter and discuss to what extent the 
coefficients of $\nu$SMEFT 
are constrained by current data in this regime.

%-----------------------------------
\section{Constraints on bosonic operators}
\label{sec:bosonic}
%-----------------------------------
While most of our study focusses on constraints on four-fermion operators,
(see section~\ref{sec:searches}), in this section we want to summarise 
limits on the bosonic operators in Tab.~\ref{tab:basis}.

For convenience, we will trade the operators $\mathcal{O}_{NB}$ and 
$\mathcal{O}_{NW}$ by $\mathcal{O}_{NA} = 
c_W\mathcal{O}_{NB}+s_W\mathcal{O}_{NW}$ and 
$\mathcal{O}_{NZ}=-s_W\mathcal{O}_{NB}+c_W\mathcal{O}_{NW}$. As we anticipated 
above, $\alpha_{NA}=c_W\alpha_{NB}+s_W\alpha_{NW}$, while 
$\alpha_{NZ}=c_W\alpha_{NW}-s_W\alpha_{NB}$. We also have the relation 
$\alpha_{NW}=\alpha_{NB}t_W$ with $t_W=s_W/c_W$.

The operator $\ope_{NZ}$ triggers the decay $Z \to \nu N$. Accounting for the 
three RH neutrinos, we find 
\begin{equation}
\Gamma (Z \to \nu N) = \frac{3 m_Z^3 v^2}{12 \pi \Lambda^4} \, 
\left[ 2(\alpha_{NZ}^{\ell})^2 + (\alpha_{NZ}^{\tau})^2 \right]\, .
\end{equation}
This process leads to the signal $Z\to \nu\nu\gamma$. The corresponding 
branching ratio is experimentally bounded 
to $\mathcal{B}(Z\to \nu \nu \gamma)<3.2\times 10^{-6}$~\cite{Acciarri:1997im}. Using that the total $Z$ width 
is $\sim 2.5$ 
GeV~\cite{Tanabashi:2018oca}, we obtain the bounds $|\alpha_{NZ}^{\ell}|< 0.37$ 
and $|\alpha_{NZ}^{\tau}|< 0.52$
for $\Lambda=1$ 
TeV. 

$Z$ decays are also triggered by $\mathcal{O}_{HN}$
\begin{equation}
\Gamma (Z \to N N) = \frac{m_Z^3 v^2}{8 \pi \Lambda^4} \, \alpha_{HN}^2 \, ,
\end{equation}
which leads to $Z\to\nu\nu\gamma\gamma$. The experimental upper bound on the 
corresponding branching
ratio is $\mathcal{B}(Z\to\nu\nu\gamma\gamma) < 3.1\times 
10^{-6}$~\cite{Tanabashi:2018oca}. 
This translates into a bound on $|\alpha_{HN}| < 0.065$ for $\Lambda = 1$ TeV.

The coefficient of the operator $\ope_{HNe}$ can be constrained by 
measurements of the $W$~boson width:
\begin{equation}
\Gamma (W \to \ell N ) = \frac{3m_W^3 v^2}{48 \pi \Lambda^4} 
\bigg\lbrace 2(\alpha_{HNe}^\ell)^2 + (\alpha_{HNe}^{\tau})^2 + s_W^2 
\bigg[2(\alpha_{NA}^\ell)^2 +(\alpha_{NA}^\tau)^2 \bigg]  \bigg\rbrace 
+\cdots\, ,
\end{equation}
where the ellipsis involve terms proportional to the very constrained 
$\alpha_{NZ}$.

To 
the best of our knowledge, there is no measurement of this branching ratio, 
while the bounds on $\alpha_{HNe}$ and $\alpha_{NA}$ from the measurement of 
the total $W$ width are very weak~\footnote{Moreover, although the 
operator $\mathcal{O}_{HNe}$  renormalises
the very much constrained $\mathcal{O}_{HN}$, the mixing is Yukawa suppressed; see Ref.~\cite{Chala:2020pbn} for
the one-loop running of all Higgs operators.}.
The best bounds on $\alpha_{NA}$ were 
actually obtained in Ref.~\cite{Butterworth:2019iff} using LHC searches for one 
photon 
and missing energy. Given our flavour assumptions, this is about 
$\alpha_{NA}^{\ell}/\Lambda^2<0.3$ TeV$^{-2}$, and hence consistent with our 
range for $m_N$; see 
section~\ref{sec:parameter_space}.
%

%-----------------------------------
\section{Searches limiting four-fermion operators}
\label{sec:searches}
%-----------------------------------
The four-fermion operators listed in Tab.~\ref{tab:basis} can have observable 
consequences for searches at $pp$ as well as $e^+e^-$ colliders. 
In the following, we will use LHC and LEP searches as well as $\tau$ and $\pi$~decay
measurements to constrain the coefficients of the $\nu$SMEFT Lagrangian. 
As a first overview, we list the considered experiments and the coefficients 
which they are sensitive to in Tab.~\ref{tab:proc_constraints}. We generally 
neglect contributions from bosonic operators, since the 
processes we consider in the following will not provide competitive bounds on bosonic operators compared to the ones presented in section~\ref{sec:bosonic}. 
Furthermore, within their bounds the bosonic operators will not have a meaningful impact on the derivation of the limits on four-fermion operators
in multi-parameter fits, so we can safely neglect them.

\begin{table}[t]
\centering
\begin{tabular}{lcccccccc}
\toprule
 & $p p \rightarrow \ell N$ & $p p \rightarrow N N$ & $\pi \rightarrow \ell N$ & $\tau \rightarrow \ell N \nu$ &  $\tau \rightarrow \pi N$ 
& $ee \to NN$ & $ee \to \nu N$ \\
 \midrule
$\ope_{eN}$ & & & &  & &$(\ell \ell)$   \\
$\ope_{uN}$ & & $(qq)$ & & &  &    \\
$\ope_{dN}$ & & $(qq)$ & & &  &   \\
$\ope_{LN}$ & & & &  & & $(\ell \ell)$  \\
$\ope_{QN}$ & & $(qq)$ & & &  &   \\

$\ope_{LNLe}$ & & & & mult & & & mult  \\
$\ope_{LdQN}$ & $(\ell qq)$ & &   & &   & \\
$\ope_{LNQd}$ & $(\ell qq)$ & & $(\ell qq)$ & & $(\tau qq)$&  \\
$\ope_{QuNL}$ & $(qq\ell)$ & & $(qq\ell)$ & & $(qq\tau)$&   \\
$\ope_{duNe}$ & $(qq\ell)$ & & $(qq\ell)$ & &  $(qq\tau)$ &  \\

\bottomrule
\end{tabular}
\caption{Overview of the considered processes as well as the parameters of 
four-fermion operators which they constrain. The notation $(XX)$ means 
that the parameter $\alpha^{XX}$ can be bound by a given process. 
For instance, the entry $(qq)$ for $\ope_{uN}$ and $pp \to NN$ stands for the process $pp \to NN$ setting a bound on $\alpha_{uN}^{(qq)}$.
We use ``Mult'' if a process is able to constrain multiple coefficients of an operator.
}
\label{tab:proc_constraints}
\end{table}

To constrain the $\nu$SMEFT parameter space, 
we will recast existing LEP and LHC searches. 
Event generation for these studies is performed with 
\texttt{MadGraph-v2.6.7}~\cite{Alwall:2014hca} at leading order, using the 
\texttt{NNPDF30\_nlo\_as\_0118} PDFs from the 
\texttt{LHAPDF} set~\cite{Buckley:2014ana} for proton collisions. We use the default \texttt{MadGraph} dynamical renormalization and factorization scales.
Parton showering, fragmentation and hadronization is performed with \texttt{Pythia v8}~\cite{Sjostrand:2014zea}. 
To recast the cuts employed in the experimental analyses, we use routines from \texttt{HepMC v2}~\cite{Dobbs:2001ck} and \texttt{Fastjet v3}~\cite{Cacciari:2011ma}. 
Detector effects are generally neglected, but we include factors to account for the detector 
efficiencies. 

Some of the considered analyses allow for jets in the final state. 
We have explicitly checked that generating our signal process with 
additional hard jets does not significantly increase the number of events. 

\subsection{LHC searches for one lepton plus one photon plus missing transverse energy}
\label{sec:lepton_photon_etmiss}
Operators which generate four-point interactions 
of two light quarks, a lepton and a RH neutrino, $\ope_{duNe}$, $\ope_{LdQN}$, $\ope_{LNQd}$ and $\ope_{QuNL}$,  contribute to the  $pp\to\ell N$ process, 
where $N$ can be any of the three RH neutrinos. 
After the decay of the RH neutrino this leads to an $\ell \gamma \etmiss$ 
signature. 
We neglect contributions from the bosonic operators $\ope_{NW}$ 
and $\ope_{HNe}$.

Given the different helicity structures of the dimension-six operators involved in this process, only the operators 
$\ope_{LdQN}$ and $\ope_{LNQd}$ interfere with each other. 
We can therefore express the number of events in different signal regions as
(compare also Eqs.~(2.1) and (2.2) of Ref.~\cite{Alcaide:2019pnf})
\begin{equation}
\begin{split}
N = \frac{1}{\Lambda^4} \bigg\{
	&\left[ (\alpha_{QuNL}^{qq\ell})^2 + 4 (\alpha_{duNe}^{qq\ell})^2 + (\alpha_{LNQd}^{\ell qq})^2 \right] {\cal A}_1 
	+ \left[ 4 (\alpha_{duNe}^{qq\ell})^2 + (\alpha_{LdQN}^{\ell qq})^2 \right] {\cal A}_2  \\
	&+2\left[ 4 (\alpha_{duNe}^{qq\ell})^2 - \alpha_{LNQd}^{\ell qq} \, \alpha_{LdQN}^{\ell qq}  \right] {\cal A}_3 
	\bigg\} \, .
\end{split}
\label{eq:coefs_Nl}
\end{equation}
CMS has carried out a search for the one lepton ($e$ or $\mu$) plus one photon plus large missing energy (and jets) signature 
based on $35.9\ifb$ of data collected at $13$~TeV  in Ref.~\cite{Sirunyan:2018psa}.
The search demands at least one photon with $p_T^\gamma > 35\gev$ and at least one lepton with $p_T^\ell> 25 \gev$. 
Signal events are required to fulfil $\etmiss > 120$~GeV and $m_T>100$~GeV
\footnote{The variable $m_T$ is defined as  
$m_T = \sqrt{2 p_T^\ell p_T^\text{miss} \left[1 - \cos (\Delta\phi (\ell, \vec{p_T}^\text{miss} )) \right]}$.} 
and are classified into different signal regions according to their $H_T$, $p_T^\gamma$ and $\etmiss$.
As we do not expect any jets in our signal final state, we concentrate on the lowest-$H_T$ signal regions, \textit{i.e.}~we consider $H_T < 100\, \gev$ only. 
We also neglect overflow bins in which an EFT description would not be valid for 
$\Lambda\sim \mathcal{O}( 1)$ TeV. Since the momentum flow through the 
four-fermion vertex can reach $\sqrt{\hat{s}} \gtrsim 1 \tev$, the validity of 
an EFT expansion is limited by the choice of the new physics scale $\Lambda$. 
We find that for the LHC processes studied in this and the next section, only 
$1\%$ of the events exceed a momentum flow of $4\tev$ through the four-fermion 
vertex. Therefore, we will present our limits for $\Lambda = 4 \tev$ in this and 
the following sections.
The definition of the four remaining signal regions
in terms of $p_T^\gamma$ and $\etmiss$ is presented in Tab.~\ref{tab:cms_Nl}, 
along with the number of data events in each region as well as the SM prediction including its uncertainty. The numerical values are directly taken from Fig.~5 of Ref.~\cite{Sirunyan:2018psa}, as a \texttt{HepData} entry for this analysis was not available.
%\enlargethispage{.5cm}

%
\begin{table}[tbh]
\centering
\begin{tabular}{lcccc}
\toprule
              & \multicolumn{2}{c}{$p_T^\gamma < 200 \,\gev$} & \multicolumn{2}{c}{$p_T^\gamma > 200 \,\gev$} \\
$\etmiss$ [GeV] & $<200$ & $[200,\,400]$  & $<200$ & $[200,\,400]$     \\          
\midrule
${\cal A}_1$   &  $3140$ & $5440$ & $1700$ & $3780$\\
${\cal A}_2$  & $1160$ & $1910$ & $590$ & $1220$\\
${\cal A}_3$  & $- 1740$ & $-2990$ & $-930$ & $-2000$ \\
\midrule
%SM &  $174 \pm 19$ ($336 \pm 44$) & $18.2 \pm 2.8$ ($27.6 \pm 4.3$)  
%& $6.5\pm4.3$ ($6.6\pm 2.4$) & $4.7\pm2.9$ ($5.0\pm 1.8$)\\
$\text{SM}_{e \gamma}$  &  $174 \pm 19$ & $18.2 \pm 2.8$  & $6.5\pm4.3$ & $4.7\pm2.9$\\
$\text{SM}_{\mu \gamma}$  &  ($336 \pm 44$) & ($27.6 \pm 4.3$)  & ($6.6\pm 2.4$) & ($5.0\pm 1.8$)\\
data & $150$ ($305$) & $32$ ($31$) & $10$ ($12$) & $6$ ($4$) \\
$s_\text{max}$ & $31.$ ($66.2$) & $26.2$ ($18.0$) & $12.5$ ($14.0$) & $8.9$ ($6.8$) \\ 
\bottomrule
\end{tabular}
\caption{CMS lepton plus photon plus missing energy: Number of expected SM events and data  in different bins of the $\etmiss$ 
	distribution in Fig.~5 of~\cite{Sirunyan:2018psa} for the $e\gamma$ ($\mu \gamma$) case, excluding the overflow bins. }
\label{tab:cms_Nl}
\end{table}

The numerical values for the coefficients ${\cal A}_i$ of Eq.~\eqref{eq:coefs_Nl},
which represent the different beyond the SM (BSM) contributions to the signal region, are also presented in
Tab.~\ref{tab:cms_Nl}. They include a 
factor of~$0.59$ to account for detector effects.
\footnote{To validate our analysis, we have used the ATLAS $8\tev$ search for heavy resonances decaying to $V\gamma$ in Ref.~\cite{Aad:2014fha}, which applies very similar selection cuts as the ones considered in the
CMS analysis~\cite{Sirunyan:2018psa} for the $W\gamma$ region. We can reproduce the event numbers in 
each bin of the $m_T^{\ell \nu \gamma}$ in Fig.~1 of Ref.~\cite{Aad:2014fha} within $20\%$. 
We did not validate on the $V\gamma$ contribution to our signal regions 
directly, because of the large number of 
correction factors applied on this background in the CMS analysis. 
These factors are not present in the ATLAS search which facilitates the validation.}

Using the information in Tab.~\ref{tab:cms_Nl}, we set limits on the relevant coefficients of 
$\alpha_{duNe}$, $\alpha_{LdQN}$, $\alpha_{LNQd}$, $\alpha_{QuNL}$ in one-parameter fits. 
We allow the BSM contribution to produce $s_\text{max}$ 
events, where $s_\text{max}$ is the maximum number of allowed additional signal events, 
determined using the CL$_s$ technique~\cite{Read:2002hq}, that we quote in Tab.~\ref{tab:cms_Nl} too.
Assuming a new physics (NP) scale of $\Lambda = 4\tev$, 
the resulting one-parameter fit limits in the electron (muon) channel are 
$|\alpha_{duNe}^{qq\ell}|<0.75 \, (0.66)$, $|\alpha_{LdQN}^{\ell qq}|<1.4 \, (1.2)$, $|\alpha_{LNQd}^{\ell qq}|, \,|\alpha_{QuNL}^{qq\ell}| < 0.78 \, (0.67)$. 
These limits come from the last bin of Tab.~\ref{tab:cms_Nl} only, 
$p_T^\gamma>200 \gev$ and $\etmiss \in [200, \, 400] \gev$, 
where there is a small underfluctuation in the muon data. 
The limits from the muon channel are hence stronger than those
from the electron channel.
For the limits on $\alpha_{LdQN}$ and $\alpha_{LNQd}$, we should take into account that the corresponding operators 
have a negative interference. We therefore marginalise over $\alpha_{LdQN}$ when constraining $\alpha_{LNQd}$ 
(and vice versa). The marginalization weakens the limits on these operators 
to $|\alpha_{LdQN}^{\ell qq}|< 3.7 \, (3.2)$, 
$|\alpha_{LNQd}^{\ell qq}| < 1.9 \,(1.6)$ in the electron (muon) channel.
Limits for lower values of $\Lambda$ could be obtained by imposing a cut on the 
momentum flow 
through the four-fermion vertex. As an example, more than $50\%$ of the events have a momentum flow through the four-fermion vertex of less than $1\tev$. This allows us to set a limit of, for instance, 
$|\alpha_{duNe}^{qq\ell}|< \sqrt{2} \cdot 0.66 \cdot (1 \tev)^2/(4\tev)^2 = 0.058$ for $\Lambda = 1 \tev$  in the muon channel.

In principle, the presented CMS search is also sensitive to $pp \to t t , \, t \rightarrow b \ell N$, 
as it allows for jets in the final state. 
However, the resulting limits are very weak, $\alpha/\Lambda^2 \sim \ope (50 )\, \tev^{-2}$.

% %-------------------------------------------------
\subsection{LHC searches for two photons plus missing transverse energy}
\label{sec:two_photons_etmiss}
Four-fermion operators containing two light quarks and two RH neutrinos can contribute to a diphoton plus 
missing energy signature at the LHC via the process $pp\to N N  \to \gamma \gamma \nu \nu $, where the $NN$ can be any pair of the three RH neutrinos 
$NN = N_1 N_1 + N_2 N_2 + N_3 N_3$.
The operators contributing to the considered signature are 
$\ope_{uN}$, $\ope_{dN}$, $\ope_{QN}$. 
The interference between either of the operators $\ope_{uN}$ and $\ope_{dN}$ with the operator $\ope_{QN}$ is helicity suppressed. 
We can therefore parametrise the number of events in different signal regions as 
\begin{align}
	N = \frac{1}{\Lambda^4} \left[
			(\alpha^{qq}_{uN})^2 \, {\cal C}_1 + (\alpha^{qq}_{dN})^2 \, {\cal C}_2 + (\alpha^{qq}_{QN})^2 \, {\cal C}_3
			\right] \, .
\label{eq:coefs_NN}
\end{align}

CMS has carried out a search for two photons plus missing energy in Ref.~\cite{Sirunyan:2019mbp}, 
based on $35.9\ifb$ of data at $\sqrt{s}=13\tev$. 
The main signal specifications are two photons with $p_T^\gamma>40\gev$ in the central detector region
 $|\eta^\gamma|<1.44$ and a significant amount of missing transverse energy $\etmiss>100\gev$.
 The analysis further vetoes events with leptons with $p_T^\ell > 25\gev$ and the two photons are 
 required to have an invariant mass $m_{\gamma\gamma} >105\gev$ and to be separated 
 by $\Delta R > 0.6$. The predicted number of SM events as well as the observed data in different $\etmiss$
 bins as provided in Tab.~2 of Ref.~\cite{Sirunyan:2019mbp} is given in Tab.~\ref{tab:cms_NN}, together with the values of $s_{\rm max}$.
\begin{table}[tbh]
\centering
\begin{tabular}{lccccc}
\toprule
$\etmiss$ [GeV] & $[100,\,115]$ & $[115,\,130]$ & $[130,\, 150]$     & $[150,\, 185]$ & $[185,\,250]$     \\          
\midrule
${\cal C}_1$  & $1090$ & $990$ & $1260$ & $1900$ & $2830$ \\
${\cal C}_2$ & $670$ & $650$ & $750$ & $1160$ & $1590$\\
${\cal C}_3$ & $1710$ & $1640$ & $2040$ & $2940$ & $4210$\\
\midrule
SM &  $110.1 \pm 7.4$ & $41.5 \pm 3.9$ & $25.9\pm3.1$ & $18.1\pm2.6$ & $10.9\pm1.8$ \\
data & $105$ & $39$ & $21$ & $21$ & $11$ \\
$s_\text{max}$ & $23.5$ & $15.0$ & $10.5$ & $14.3$ & $9.6$ \\ 
\bottomrule
\end{tabular}
\caption{CMS diphoton plus missing energy: Number of expected events and data in different bins of the $\etmiss$ 
	distribution in Tab.~3 (post-fit) of Ref.~\cite{Sirunyan:2019mbp}. 
	We neglect the overflow bin. }
\label{tab:cms_NN}
\end{table}

We validate our implementation of the CMS signal region definition using the $Z\gamma\gamma$ background. 
This background is subdominant, accounting for between $1\%$ and $20\%$ 
of the total background only (depending on the missing energy bin). 
For the analysis validation, however, it has the advantage that 
it comes purely from Monte Carlo simulation and no data-driven correction 
factors were applied. 
Moreover, this background is the only one with a genuine $\gamma\gamma E_T^{miss}$ signature from $\nu\nu\gamma\gamma$ 
and therefore the detector effects relevant for it will best represent the detector effects on our signal. 
We find that using a global 
scale factor of $0.59$ to account for detector efficiencies, we can reproduce the $\etmiss$ 
distribution of the $Z\gamma\gamma$ background within $5\%$. 

Recasting the CMS analysis for our heavy neutrino pair production signal, we find the 
parametrization of the event numbers in different $\etmiss$ bins in terms of 
the parameters~${\cal C}_i$ of Eq.~\eqref{eq:coefs_NN} displayed in 
Tab.~\ref{tab:cms_NN}. 
Translating the parametrization into limits on the $\nu$SMEFT coefficients, we observe that the highest bin
provides the best sensitivity, as expected. The resulting limits from the last bin only
are 
$|\alpha_{uN}^{qq}| < 0.93 $, $|\alpha_{dN}^{qq}|<1.2$, $|\alpha_{QN}^{qq}|<0.77$, where we 
again use $\Lambda = 4 \tev$ to stay within the range of validity of the EFT 
description. 
As expected from the PDFs, the limits on up-quark couplings to $NN$ are stronger than the ones 
from down-quark couplings and the strongest limits arise for the operator influencing both up-quark
and down-quark couplings. 
Limits for lower new physics scales $\Lambda$ could be obtained by imposing a cut on the momentum 
flow through the four-fermion vertex. About $40 \%$ of all events 
have a momentum flow of $\sqrt{\hat{s}} < 1 \tev$, for $\sqrt{\hat{s}} < 2 \tev$ 
the fraction rises up to $80\%$. Taking these values into account, one can 
rescale the above limits for lower values of $\Lambda$.
%\enlargethispage{.5cm}

%-----------------------------------------------------------------------
\subsection{Pion decays}
\label{sec:pion}

Given low RH neutrino masses $m_N \lesssim m_\pi$, operators which generate four-point interactions 
of two light quarks, a lepton and a RH neutrino, $\ope_{duNe}$, $\ope_{LdQN}$, $\ope_{LNQd}$ and $\ope_{QuNL}$, 
do not only contribute to the $pp\to\ell N$ process as discussed in section~\ref{sec:lepton_photon_etmiss}, but
they also trigger the pion decay $\pi \to \ell \gamma \nu$. 
In the following, we will neglect the operator~$\ope_{LdQN}$, for which the pion form factor is hard to estimate.

The pion decay width, including all fermion masses and a factor of $3$ to account for
the three RH neutrino flavours, is described by 
(see also Ref.~\cite{Carpentier:2010ue})
\begin{eqnarray}
\begin{split}
\Gamma(\pi \rightarrow \ell N)	=& \frac{3 f_\pi^2 k}{16\pi m_\pi^2 \Lambda^4} \, \Bigg\{  
				 \alpha_V^2 
				\left[ (m_\ell^2 + m_N^2) (m_\pi^2 -m_N^2 - m_\ell^2) 
				   + 4 m_\ell^2 m_N^2 \right] \\
					& \hspace*{1.9cm} + 2 \, \alpha_V \, \alpha_P \,
					m_\ell \, \frac{m_\pi^2}{m_u+m_d} \, \left[ m_\pi^2 + m_N^2 - m_\ell^2 \right] \\
					& \hspace*{1.9cm} + \alpha_P^2 \, \left( \frac{m_\pi^2}{m_u+m_d} \right)^2 \,
					  (m_\pi^2 -m_N^2 - m_\ell^2)  \Bigg\} \, ,
\end{split}		
\label{eq:width_pion}	
\end{eqnarray}
with $f_\pi \sim 131 \mev$. Here, $\alpha_V= \alpha_{duNe}^{qq\ell}$ and 
$\alpha_P =  ( \alpha_{QuNL}^{qq \ell} -  \alpha_{LNQd}^{\ell  qq} ) $ denote the contributions from 
operators with vector couplings and pseudo-scalar couplings respectively, 
and $k$ is the magnitude of the three-momenta of the lepton and neutrino
in the center-of-mass (c.o.m) frame
\footnote{We display the $m_N$ dependence of Eq.~\eqref{eq:width_pion} in 
Appendix~\ref{app:mass_dep}, Fig.~\ref{fig:pionDecay_mN}.}:
\begin{equation}
 k = \frac{1}{2m_\pi} \sqrt{\left(m_\pi^2 - (m_l +m_N)^2 \right) \, \left(m_\pi^2 - (m_l - m_N)^2 \right) }
   \,  \overset{m_N = 0}{=} \, \frac{m_\pi^2 - m_\ell^2}{2m_\pi} \, .
\end{equation}
Due to the helicity suppression for vector couplings, $\alpha_{duNe}^{qq\ell} $ will be much less constrained than
the combination $(\alpha_{QuNL}^{qq\ell} -  \alpha_{LNQd}^{\ell qq})$.

We compare the BSM pion decay width to the experimental measurements 
of the $\pi \to \ell \gamma \nu$~branching ratio which we list in Tab.~\ref{tab:exp_pionDecays}.   
We also cite the corresponding theory prediction and determine
the maximally allowed BSM contribution to the decay width~$\Delta \Gamma^\text{BSM}$ 
which we define as 
\begin{equation}
\Delta\Gamma^\text{BSM} = \left \{~~ \begin{matrix} \, \Gamma_\text{exp} - \Gamma_\text{theo} + 2 \sigma_{\Gamma_\text{exp}} \\
		   2 \sigma_{\Gamma_\text{exp}} \, , ~\text{where no measurement/prediction is available.} \end{matrix} \right .
\label{eq:DGamma_tau}
\end{equation}

%--------------------
\begin{table}[tbh]
%\vspace*{-7mm}
\centering
\begin{tabular}{lcc}
\toprule
  & $\pi \to e \gamma \nu$ & $\pi \to \mu \gamma \nu$  \\
BSM process & $\pi \to e  N$ & $\pi \to \mu  N$  \\  
 \midrule
 $\text{BR}_\text{exp}$ & $(7.39 \pm 0.05 ) \times 10^{-7}$~\cite{Bychkov:2008ws} 
 	& $(2.00 \pm 0.25 ) \times 10^{-4}$~\cite{Bressi:1997gs}  \\
  $\text{BR}_\text{theo}$  & $7.411 \times 10^{-7}$ & $2.283 \times 10^{-4}$  \\
  % cuts & $E_\gamma> 10 \mev$, $\theta_{e\gamma}>40^{\circ}$ & $E_\gamma > 1 \mev$\\
  \midrule
   $\Gamma_\text{exp}$ [GeV] & $(186.7 \pm 1.3) \times 10^{-25}$ & $(50.5 \pm 6.3) \times 10^{-22}$ \\
   $\Delta \Gamma^\text{BSM}$ [GeV] & $2.0 \times 10^{-25}$ & $5.5 \times 10^{-22}$  \\
 \bottomrule
\end{tabular}
\caption{Measured values and theoretical predictions of branching ratios for different $\pi$ decays. The theory predictions are taken from the corresponding experimental 
references. 
For convenience, we also translate the measured branching ratio into decay width 
and list the allowed contribution from BSM processes according to Eq.~\eqref{eq:DGamma_tau}.}
\label{tab:exp_pionDecays}
\end{table}
%--------------------

The resulting bounds in a one-parameter fit in the limit $m_N=0$ are 
$|\alpha_{duNe}^{qq\ell}| < 73 \, (0.04)$, 
and $|\alpha_{QuNL}^{qq\ell} - \alpha_{LNQd}^{\ell qq}| < 1.3 \times 10^{-5}$ ($0.002$) in the 
electron (muon) channel respectively, for $\Lambda = 1\,\tev$.
As  $\alpha_{QuNL}^{qq\ell}$ and  $\alpha_{LNQd}^{\ell qq}$ interfere negatively, we should 
note, however, that any BSM contributions for one operator can in principle be 
cancelled exactly by the other one. We can only constrain the difference between 
the two coefficients, but each individual coefficient is unconstrained. 
For $m_N=0.1 \gev$, only the electron decay channel is kinematically open. The limits
from this channel are $|\alpha_{duNe}^{qq \ell}|  <7.7\times 10^{-4}$ and 
$|\alpha_{QuNL}^{qq \ell} -\alpha_{LNQd}^{\ell qq}| < 2.7\times 10^{-5}$.

%-----------------------------------------------------------------------
\vspace*{-1mm}
\subsection{Tau decays}
\label{sec:tau}
%-----------------------------------------------------------------------
The four-fermion operators in Tab.~\ref{tab:basis} can contribute to 
$\tau$~decays with 
a photon in the $\tau \to \ell \nu N \to \ell \nu \nu \gamma$
and 
$\tau \to \pi N \to \pi \gamma \nu$ channels. Each of these processes 
is sensitive to a different set of operator coefficients; see 
Tab.~\ref{tab:proc_constraints}. 
In Tab.~\ref{tab:exp_tauDecays}, we list the experimentally measured branching 
ratios of the considered decay channels along with their theory prediction. 
\enlargethispage{.5cm}
%--------------------
\begin{table}[htb]
\centering
\begin{tabular}{lccc}
\toprule
  & $\tau \to e \gamma \nu$ & $\tau \to \mu \gamma \nu$ & $\tau \to \pi \gamma \nu$ \\
BSM process & $\tau \to e \nu N$ & $\tau \to \mu \nu N$  & $\tau \to \pi N$ \\  
 \midrule
 $\text{BR}_\text{exp}$ & $(1.83 \pm 0.05 )\,\%$ & $(0.367 \pm 0.008)\, \%$ & $( 3.8 \pm 1.5 )\times 10^{-4}$\\
  $\text{BR}_\text{theo}$  & $1.645 \,\%$~\cite{Fael:2015gua} & $0.3572\,\%$~\cite{Fael:2015gua}  &\\
%  cuts  & $E_\gamma^*> 10 \mev$ & $E_\gamma^*> 10 \mev$ &\\
  \midrule
   $\Gamma_\text{exp}$ [GeV] & $(83. \pm 2.) \times 10^{-16}$ & $(41. \pm 2.) \times 10^{-15} $ & $\pm 3.4 \times 10^{-16}$ \\
   $\Delta \Gamma^\text{BSM}$ [GeV] & $6.5 \times 10^{-15}$ & $5.9\times 10^{-16}$ & $6.8 \times 10^{-16}$ \\
 \bottomrule
\end{tabular}
\caption{Measured values and theoretical predictions of branching ratios for different $\tau$ decays.
All experimental values are taken from the PDG~\cite{Tanabashi:2018oca}. 
For convenience, we also translate the measured branching ratio into decay width 
and list the allowed contribution from BSM processes according to Eq.~\eqref{eq:DGamma_tau}.
Note the discrepancy between theory and experiment in the $\tau \to e \nu N$ channel.
}
\label{tab:exp_tauDecays}
\end{table}
%--------------------

\vspace*{-1mm}
\subsubsection*{Tau decays to $\tau \to \ell \nu N$}
The $\tau \to \ell \nu N$ process is sensitive to different coefficients of the operator~$\ope_{LNLe}$.
Neglecting the mass of the RH neutrino, they contribute to the decay width through 
\begin{equation}
\begin{split}
\Gamma (\tau \rightarrow \ell_i N \nu_i) =& \frac{m_\tau^5}{2048 \pi^3 \Lambda^4} (\alpha_{LNLe}^{\ell \ell \tau})^2\,\, \\
\Gamma (\tau \rightarrow \ell_i \bar{\nu_i} N) =& \frac{m_\tau^5}{2048 \pi^3 \Lambda^4} 
				\left[ (\alpha_{LNLe}^{\ell\tau\ell})^2 + (\alpha_{LNLe}^{\tau\ell\ell})^2 - \alpha_{LNLe}^{\ell\tau\ell} \alpha_{LNLe}^{\tau\ell\ell} \right]~, \\
\Gamma (\tau \rightarrow \ell_i N \nu_\tau)  =& \frac{m_\tau^5}{2048 \pi^3 \Lambda^4} 
				\left[ (\alpha_{LNLe}^{\ell\tau\tau})^2 + (\alpha_{LNLe}^{\tau\ell\tau})^2 - \alpha_{LNLe}^{\ell\tau\tau} \alpha_{LNLe}^{\tau\ell\tau} \right]~, \\
\Gamma (\tau \rightarrow \ell_i \bar{\nu_\tau} N)  =& \frac{m_\tau^5}{2048 \pi^3 \Lambda^4} (\alpha_{LNLe}^{\tau\tau\ell})^2\,.
\end{split}
\end{equation}

The correction factor to include the mass of the RH neutrinos reads
\footnote{We display the $m_N$ dependence of Eq.~\eqref{eq:width_tau_mN} in 
Appendix~\ref{app:mass_dep}, in the left panel of Fig.~\ref{fig:tauDecay_mN}.}
\begin{equation}
 \frac{\Gamma (x = \frac{m_N}{m_\pi})}{\Gamma (m_N=0)} = 1 - 8 x^2 + 8 x^6 - x^8 - 24 x^4 \log x\,.
\label{eq:width_tau_mN}
\end{equation}
We set limits on the components of $\alpha_{LNLe}$ by letting 
the EFT contribution account for $\Delta \Gamma^\text{BSM}$ as given in Eq.~\eqref{eq:DGamma_tau}. 
Assuming that only one of the $\alpha_{LNLe}$ components is non-zero, 
we can set a limit of $|\alpha_{LNLe}| < 4.9$ ($1.5$) for the decay into an electron (muon) and 
relatively light RH neutrinos $m_N=0.1\, \gev$.
The difference between the limits in the electron and muon channels results entirely from 
$\Delta \Gamma^\text{BSM}_{\tau \to e \gamma \nu} > \Delta \Gamma^\text{BSM}_{\tau \to \mu \gamma \nu}$; 
we do not include the mass of the charged leptons in our calculations. 
The obtained limits for $m_N=0.1 \,\gev$ are already quite weak 
and are further diluted when we consider heavier RH neutrinos; see the left panel 
of Fig.~\ref{fig:tauDecay_limits}. At $m_N = 1\, \gev$, for instance, the limits become
$\alpha_{LNLe} < 16$ ($4.8$) in the electron (muon) channel, respectively, assuming $\Lambda = 1 \tev$. 

For those components with negative interferences, we marginalise over 
the other relevant components when setting limits. 
The obtained bounds in the muon decay channel for $m_N = 0.1 \gev$ are
$|\alpha_{LNLe}^{\ell\ell\tau}| , \, |\alpha_{LNLe}^{\tau\tau\ell}|< 1.5$ (no interference) and 
$|\alpha_{LNLe}^{\ell \tau \ell}| , \, |\alpha_{LNLe}^{\tau \ell \ell}| , \,$ $|\alpha_{LNLe}^{\ell \tau\tau}| , \, |\alpha_{LNLe}^{\tau\tau\ell }|  < 1.7$.

\begin{figure}
	\centering
	\includegraphics[width=.45\textwidth]{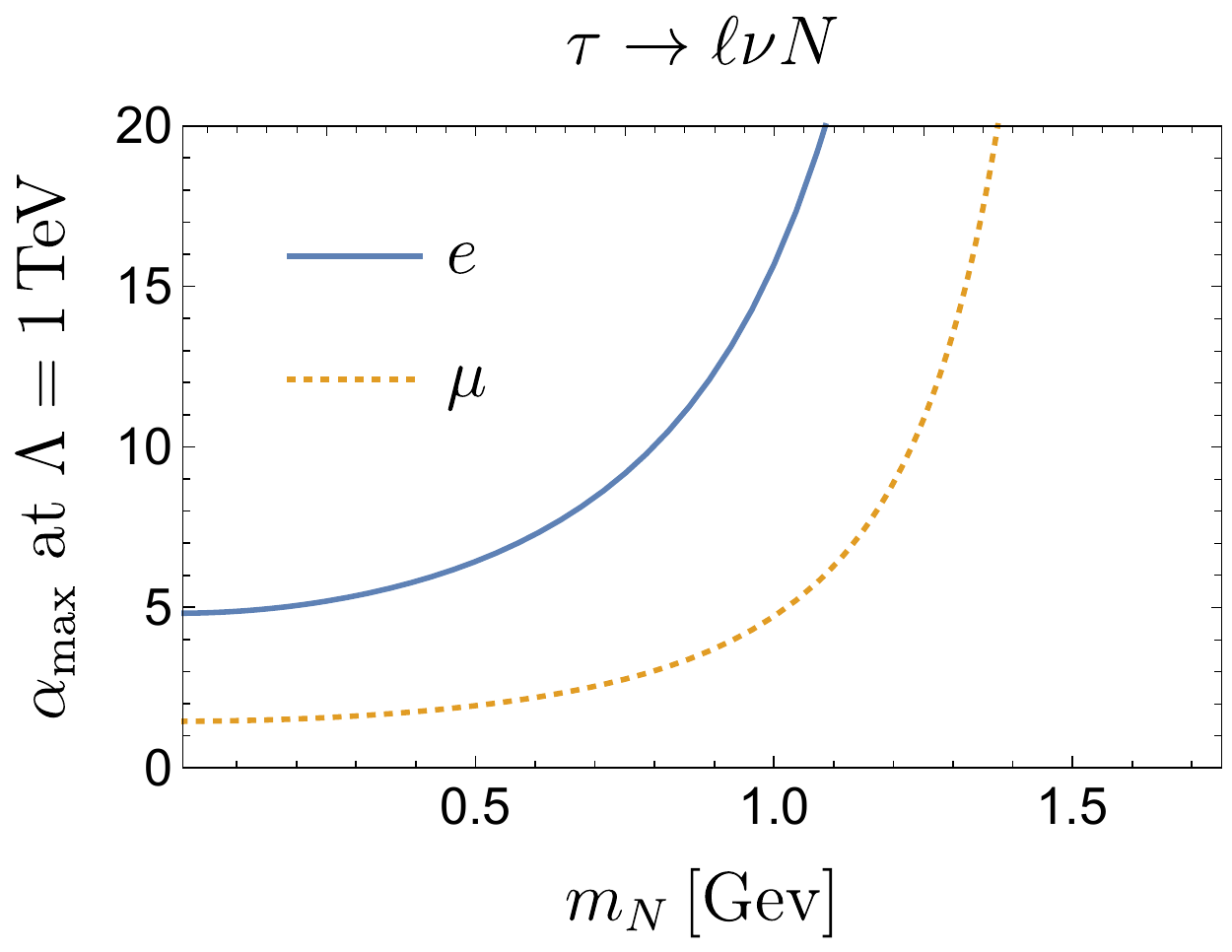}
	\quad
	\includegraphics[width=.45\textwidth]{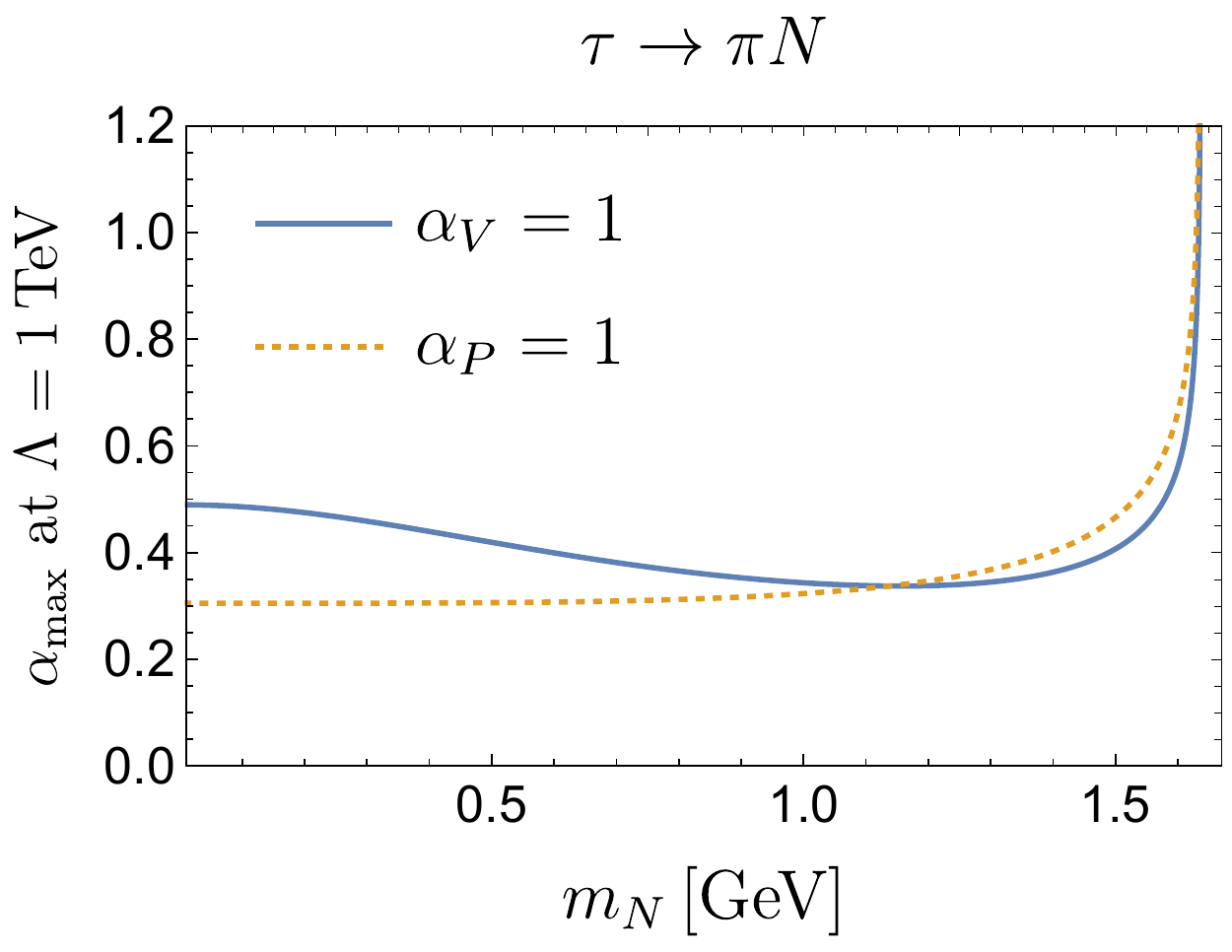}
	\caption{RH neutrino mass dependence of the one-parameter fit limits 
	from $\tau$~decays. Left: limits on $\alpha_{LNLe}$ 
	from the process $\tau \rightarrow \ell \nu N$. 
	The mass on the charged leptons has been neglected.
	Right: limits on $\alpha_V= \alpha_{duNe}^{qq\tau} $
	 and $\alpha_P =  ( \alpha_{QuNL}^{qq\tau} -  \alpha_{LNQd}^{\tau qq} ) $ 
	 from $\tau \to \pi N$.
	}
	\label{fig:tauDecay_limits}
\end{figure}

\subsubsection*{Tau decays to $\tau \to \pi N$}

The same operators that contribute to leptonic pion decays (see section~\ref{sec:pion}) will 
also add to $\tau$~decays to pions (with different coefficients). 
We can therefore use the search for 
$\tau$~decays in the $\tau \rightarrow \pi \gamma \nu$ channel, to 
constrain the coefficients $ \alpha_{duNe}^{qq\tau}$, 
$\alpha_{QuNL}^{qq\tau}$ and $ \alpha_{LNQd}^{\tau qq}$. 
The observed branching ratio in this channel is given in Tab.~\ref{tab:exp_tauDecays}.

The decay width of the $\tau$~lepton into a pion and a RH neutrino 
is structurally very similar to the pion decay width in Eq.~\eqref{eq:width_pion}:
\begin{eqnarray}
\begin{split}
\Gamma_{\tau \rightarrow \pi N }
				=& \frac{3 f_\pi^2 k}{16\pi m_\tau^2 \Lambda^4} \, \Bigg\{  
				 \alpha_V^2 
				\left[ (m_\tau^2 + m_N^2) (m_\tau^2 +m_N^2 - m_\pi^2) 
				+ 4 m_\tau^2 m_N^2 \right] \\
					& \hspace*{1.9cm} + 2 \, \alpha_V \, \alpha_P \,
					m_N \, \frac{m_\pi^2}{m_u+m_d} \, ( 3 m_\tau^2 + m_N^2 - m_\pi^2 )
					\\
					& \hspace*{1.9cm} 
					+ \alpha_P^2 \, \left( \frac{m_\pi^2}{m_u+m_d} \right)^2 \,
					  (m_\tau^2 +m_N^2 - m_\pi^2) \Bigg\} \, ,
\end{split}		
\label{eq:width_tau_to_pion}	
\end{eqnarray}
where $\alpha_V= \alpha_{duNe}^{qq\tau}$ and 
$\alpha_P =  ( \alpha_{QuNL}^{qq\tau} -  \alpha_{LNQd}^{\tau qq} ) $ denote the contributions from 
the operators with vector couplings and pseudo-scalar couplings respectively
and we use $f_\pi \sim 131 \mev$ again for the pion form factor.
$k$ is the magnitude of the three-momenta of the pion and neutrino
in the c.o.m frame
\footnote{We display the $m_N$ dependence of Eq.~\eqref{eq:width_tau_mN2} in 
Appendix~\ref{app:mass_dep}, in the right panel of Fig.~\ref{fig:tauDecay_mN}.}
\begin{equation}
 k = \frac{1}{2m_\tau} \sqrt{\left(m_\tau^2 - (m_\pi +m_N)^2 \right) \, \left(m_\tau^2 - (m_\pi - m_N)^2 \right) }
   \,  \overset{m_N = 0}{=} \, \frac{m_\tau^2 - m_\pi^2}{2m_\tau} \, .
   \label{eq:width_tau_mN2}
\end{equation}
As expected for a two-body decay, the decay width 
does not drop as quickly with $m_N$ as for the three-body decays considered before. 

To the best of our knowledge, there is no SM estimate for the width $\Gamma_{\tau \to \pi \gamma \nu}$. 
We therefore set conservative limits on the dimension-six operators involved  by letting
the BSM contribution account for twice the uncertainty of the experimental 
measurement, see Tab.~\ref{tab:exp_tauDecays}. 
For $m_N = 0.1 \,\gev$ and $\Lambda = 1\, \tev$, we obtain limits of 
$|\alpha_{duNe}^{qq\tau}| < 0.49$ and 
$|( \alpha_{QuNL}^{qq\tau} -  \alpha_{LNQd}^{\tau qq} ) | < 0.30$. 
These limits are rather insensitive to the RH neutrino mass (see the right panel of
Fig.~\ref{fig:tauDecay_limits}), as long as $m_N < m_\tau - m_\pi$ of course.

%------------------------------------
\subsection{LEP searches for single and multiple photons}
\label{sec:LEP}
%------------------------------------
LEP searches for a single high-energy photon or multiple photons and missing 
energy can be used 
to constrain the interactions of (first family) leptons to RH as well as LH neutrinos
via the $e e \to NN \to \gamma \gamma  \nu \nu$ and $e e \to \nu N \to  \gamma  \nu \nu$ channels. 
The LEP L3 analysis for single and multi-photon events with missing energy~\cite{Achard:2003tx} provides $\mathcal{L} = 619.1 \ipb$ of data at an average c.o.m energy of $\sqrt{s} = 197.6 \gev$. 
In the following,  we will use this experimental analysis to constrain the 
coefficients $\alpha_{eN}^{\ell\ell}$, $\alpha_{LN}^{\ell\ell}$ in $ee \to NN$ and 
$\alpha_{LNLe}$  in $ee \rightarrow \nu N $.

\subsubsection*{LEP search for multiple photons and missing energy}
The coefficients $\alpha_{eN}^{\ell\ell}$, $\alpha_{LN}^{\ell\ell}$ contribute to the process 
$ee \rightarrow NN \rightarrow \gamma \gamma \nu \nu$, where again the $NN$ can be any pair of the three RH neutrinos 
$NN = N_1 N_1 + N_2 N_2 + N_3 N_3$.
The interference of the operators contributing to $ee \rightarrow NN \rightarrow \gamma \gamma \nu \nu$ is helicity suppressed and we parametrise the number of events in terms of the $\nu$SMEFT coefficients as
\begin{align}
	N = \frac{1}{\Lambda^4} \left[
			 (\alpha^{\ell \ell}_{eN})^2 + (\alpha^{\ell \ell}_{lN})^2  \right]  \, {\cal D} \, ,
\label{eq:BSM_LEP_multi}
\end{align}
where the numerical value for $\cal{D}$ is given in Tab.~\ref{tab:LEP_multiphoton}
for two overlapping signal regions.

The considered LEP L3 analysis in the multi-photon channel focuses on events with (at least) two photons with $E_\gamma > 1 \,\gev$ 
and a transverse momentum of the diphoton system of $p_T^\gamma > 0.02 \sqrt{s} \approx 4 \gev$.
The hardest photon has to be inside the range $\theta_{\gamma_1} \in [14^\circ, \, 166^\circ]$.
A cut on the acoplanarity $ | | \phi_{\gamma_1} - \phi_{\gamma_2} | - \pi |> 2.5^\circ$ severely 
reduces the sensitivity on our signal process, where the photons are mostly back-to-back. 
However, we can still deduce meaningful results from the LEP analysis. 

We base our limits on the missing mass $m_\text{miss}$ distribution in 
the LEP analysis, where $m_\text{miss}$ is defined as the invariant mass of 
the missing momentum.
Given the fact that $m_\text{miss}$  is expected to peak around 
the $Z$~boson mass $m_Z=91\, \gev$ for the SM background, we will only consider the range 
$m_\text{miss} \in [120, \, 210]\,\gev$. We analyse 
two non-exclusive signal regions: the \textit{full} region contains all events with both 
photons in the full detector region, defined in Tab.~\ref{tab:LEP_multiphoton}, 
whereas the \textit{central} region only accepts events with both 
photons in the central 
detector area.

In Tab.~\ref{tab:LEP_multiphoton}, we present the LEP data in the full and central
signal regions 
along with their SM prediction and uncertainties, 
the experimental efficiencies and the resulting 
upper $95\%$~CL limit on additional contributions to these regions.  
We also list the numerical values of the parameters~${\cal D}_i$ for the 
BSM contributions according to Eq.~\eqref{eq:BSM_LEP_multi}.
Systematic uncertainties are very small compared to the statistical ones in this analysis 
and can hence safely be neglected.
\begin{table}[thb]
\centering 
\begin{tabular}{lcc}
\toprule
& full & central\\
\midrule
$\theta_\gamma$ range & $[11^\circ, \, 169^\circ ]$ & $[43^\circ, \, 137^\circ]$ \\
average $\epsilon_\text{exp}$ & $55 \%$ & $70\%$ \\
$\mathcal{D}$ & $14.1$ & $7.9$ \\
\midrule
data & $31$ & $5$ \\
SM & $39.4 \pm 6.3$ & $10.0 \pm 3.2$\\
$s_\text{max}$ & $13.3$ & $6.9$ \\
\bottomrule
\end{tabular}
\caption{LEP data, expected number of events and corresponding CLs limit for the multi-photon selection 
  for the two considered angular ranges for the photons. We also list the considered
  average detector efficiencies. }
\label{tab:LEP_multiphoton}
\end{table}

We have validated our analysis using the SM $ee \to \gamma \gamma (\gamma) \nu \nu$ 
process. We can reproduce the total cross sections for the full 
and central regions using the respective detector efficiencies as given in Tab.~\ref{tab:LEP_multiphoton}. 

The resulting limits on the $\nu$SMEFT coefficients are $|\alpha_{eN}^{\ell \ell }|, \, |\alpha_{LN}^{\ell \ell}| < 0.97\, (0.93)$
in the full (central) detector region, for $\Lambda = 1 \tev$.
 
%------------------------------------
\subsubsection*{LEP search for a single photon and missing energy}
%------------------------------------

The coefficients $\alpha_{LNLe}^{\ell \ell \ell}$, $\alpha_{LNLe}^{\tau \ell \ell}$, $\alpha_{LNLe}^{\ell \tau \ell}$ 
can be constrained using the process 
$ee \rightarrow \nu N \rightarrow \gamma \nu \nu$, where $N$ is any of the 
three RH neutrinos. 
We parametrise the $\nu$SMEFT contributions to the number of events 
in this channel as
\begin{align}
	N = \frac{1}{\Lambda^4} \left[  (\alpha^{\ell \ell \ell}_{LNLe})^2  + (\alpha^{\tau \ell \ell}_{LNLe})^2
	+ (\alpha^{\ell \tau \ell}_{LNLe})^2 - (\alpha^{\tau \ell \ell}_{LNLe}) (\alpha^{\ell \tau \ell}_{LNLe})  \right]  \, {\cal E} \, ,
\end{align}
where the numerical value of ${\cal E}$ is listed in Tab.~\ref{tab:LEP_singlephoton}.

We now make use of the single high-energy photon events region of the
LEP L3 analysis discussed above~\cite{Achard:2003tx}.
For this signature, the analysis requires exactly one photon with $p_T^{\gamma} > 0.02 \sqrt{s} \approx 4 \gev$ in the region 
$\theta_{\gamma} \in [14^\circ, \, 166^\circ]$.

We again restrict ourselves to the 
missing mass range $m_\text{miss} \in [120, \, 210]\gev$ to exclude the main
peak of the SM background from our signal regions and list the number of events, 
the SM prediction and its uncertainty in Tab.~\ref{tab:LEP_singlephoton}. Systematic 
uncertainties can again safely be neglected. 
%-------------------
\begin{table}[thb]
\centering
\begin{tabular}{lcc}
\toprule
& full & central\\
\midrule
$\theta_\gamma$ range & $[11^\circ, \, 169^\circ ]$ & $[43^\circ, \, 137^\circ]$ \\
average $\epsilon_\text{exp}$ & $70 \%$ & $80\%$ \\
${\cal E}$ & $383.2$ & $ 331.7$\\
\midrule
data & $874$ & $533$ \\
SM & $845 \pm 29$ & $499 \pm 22$\\
$s_\text{max}$ & $105$ & $92$ \\
\bottomrule
\end{tabular}
\caption{LEP data, expected number of events and corresponding CLs limit for the single high-energy photon selection 
  for the two considered angular ranges of the photon. We also list the considered
  average detector efficiencies. }
\label{tab:LEP_singlephoton}
\end{table}
%-------------------
We have again validated our analysis using the SM background, $ee \rightarrow \gamma (\gamma) \nu \nu$. 
We can reproduce the total cross section in the full and central signal regions within $5 \%$ when 
including the detector efficiencies listed in Tab.~\ref{tab:LEP_singlephoton}.

The resulting limits on the coefficients of $\alpha_{LNLe}$ are $|\alpha_{LNLe}^{\ell \ell \ell}| < 0.52\, (0.53)$
in the full (central) detector region, assuming $\Lambda = 1 \,\tev$. For 
 $\alpha_{LNLe}^{\tau \ell \ell}$ and $\alpha_{LNLe}^{\ell \tau \ell}$, for which the corresponding 
 operators interfere negatively, the limit is diluted to 
 $|\alpha_{LNLe}^{\tau \ell \ell}|, \, |\alpha_{LNLe}^{\ell \tau \ell}| <0.60$ ($0.61$). These 
 limits are much stronger than the corresponding limits from $\tau$ decays; see section~\ref{sec:tau}. Note also that these limits are less dependent on the 
 mass of the RH neutrinos. 

\section{Limits on contact interactions}
\label{sec:limits}

Let us start our discussion of the limits on contact interactions from those which 
are independent of the RH neutrino mass. This is definitely the case 
of bounds from the LHC, 
$pp \rightarrow \ell N$ and $pp \rightarrow NN$, and LEP, $ee \to NN$ and $ee \to \nu N$, where $m_N$ is negligible 
compared to the large c.o.m energies. The only mass dependence stems from 
the RH neutrino branching ratio which we assume to be $100\%$ for the 
$N \rightarrow \gamma \nu$ channel throughout. 
At masses approaching $m_N = 10 \gev$, the branching 
ratio can be slightly reduced by new tree-level decay modes of $N$ opening up,
which we neglect in our analysis. 

For the $pp \rightarrow NN$, for which none of the contributing operators interfere, limits on the relevant coefficients
$\alpha_{uN}^{qq}$, $\alpha_{dN}^{qq}$ and $\alpha_{QN}^{qq}$ can be extracted from 
one-parameter fits. The resulting bounds are presented in Tab.~\ref{tab:lim_LHC}, where we 
assume $\Lambda = 4 \tev$ to stay within the range of validity of the EFT description. 

As already pointed out in section~\ref{sec:lepton_photon_etmiss}, in the 
$pp \rightarrow \ell N$ channel, the limits on $\alpha_{duNe}$ and  
$\alpha_{QuNL}$ can be extracted from one-parameter fits as well, as the 
corresponding operators do not interfere with any other operator contributing 
to this channel. 
For the limits on $\alpha_{LdQN}$ and $\alpha_{LNQd}$, on the other hand, 
we account for the negative interference of the  corresponding operators 
by marginalizing over one parameter when constraining the other. The resulting
limits are displayed in Tab.~\ref{tab:lim_LHC}. Since we assume $e$-$\mu$ universality, 
we only present the limits from the $\mu \gamma$ channel which gives 
the stronger constraints. 

The LEP searches for a single photon or multiple photons accompanied by missing energy provide constraints on the 
parameters $\alpha_{eN}^{\ell\ell}, \,\alpha_{LN}^{\ell\ell}$ and 
multiple coefficients of $\alpha_{LNLe}$ respectively. 
In multi-photon production, there are no negative interferences between different operators 
and we can directly copy the limits obtained in section~\ref{sec:LEP} into 
Tab.~\ref{tab:lim_RRRR}.
The fact that these bounds are more than an order 
of magnitude weaker than the bounds on the structurally similar 
operators $\alpha_{uN}^{qq}$, $\alpha_{dN}^{qq}$ and $\alpha_{QN}^{qq}$ 
from $pp \to NN$ is a result not only of lower energy at LEP, but also 
of the strong acoplanarity cut applied by LEP which reduces the 
sensitivity to BSM contributions with back-to-back photons. 
For the single high-energy photon analysis we marginalise over the contributing 
coefficients of $\alpha_{LNLe}$. 
The constraints on $\alpha_{LNLe}^{\tau \ell \ell}$ and $\alpha_{LNLe}^{\ell \tau \ell}$
are much stronger than those derived from $\tau$ decays in the $\tau \to \ell N$ channel. 

Limits from tau and pion decays are a lot more sensitive to the RH neutrino masses
than the limits from direct production at colliders discussed so far. 
However, for low RH neutrino masses, especially pion decays 
provide very strong constraints. 
For $\tau$~decays in the $\tau \to \ell \gamma \nu$ channel, 
the limits from the muon channel are stronger than 
the ones from the electron channel due to their different experimental uncertainties.
For a RH neutrino mass of $m_N \leq 0.1 \gev$ and $\Lambda = 1 \tev$, 
we can set a limit of $|\alpha_{LNLe}| < 1.5$ on those components of $\alpha_{LNLe}$
which do not interfere, \textit{i.e.}~$\alpha_{LNLe}^{\ell\ell \tau}$ and $\alpha_{LNLe}^{\tau \tau \ell}$. 
For those components with negative interferences, we marginalise over 
the relevant other components when setting limits. 
We obtain a limit of 
$|\alpha_{LNLe}^{\ell \tau \ell}| , \, |\alpha_{LNLe}^{\tau \ell \ell}| , \, |\alpha_{LNLe}^{\ell \tau\tau}| , \, |\alpha_{LNLe}^{\tau\tau\ell }|  < 1.7$.

The $\tau$~decay channel $\tau \to \pi N$ lets us constrain the coefficients 
$\alpha_{duNe}^{qq\tau}$ as well as the difference 
$|( \alpha_{QuNL}^{qq\tau} -  \alpha_{LNQd}^{\tau qq} ) |$. 
The limits, which are largely independent of $m_N$ are presented 
in Tab.~\ref{tab:lim_RRRR} and Tab.~\ref{tab:lim_LRRL}.

Limits from pion decays can only be derived for low RH neutrino masses. In the region $m_N < m_\pi$, 
however, pion decays can set strong bounds. At $m_N < 0.1 \gev$, we find limits of 
$|\alpha_{duNe}^{qq\ell}|<7.7\times 10^{-4}$ and 
$|\alpha_{QuNL}^{qq\ell} -\alpha_{LNQd}^{\ell qq}| < 2.7\times 10^{-5}$, assuming $\Lambda = 1 \tev$. 
Combining the limit on the $|\alpha_{QuNL}^{qq\ell} -\alpha_{LNQd}^{\ell qq}|$ difference with the constraints from $pp \rightarrow \ell N$, allows us to reduce the limit on $|\alpha_{LNQd}^{\ell qq}| < 0.042$.

\begin{table}[tbh]
\centering
\begin{tabular}{lccc}
\toprule
 coefficient & $\alpha_\text{max}$ for $\Lambda= 4 \tev$ & $\Lambda_\text{min}$  [TeV] for $\alpha=1$ & observable\\
 \midrule
$\alpha_{QN}^{qq}$ & $0.77$ & $4.6$ & $pp \rightarrow NN$  \\[1.5mm]

$\alpha_{uN}^{qq}$ & $0.93$ & $4.2$ & $pp \rightarrow NN$  \\[1.5mm]

$\alpha_{dN}^{qq}$ & $1.2$ & $3.6$ & $pp \rightarrow NN$ \\[1.5mm]
$\alpha_{duNe}^{qq\ell}$ & $0.66$ & $4.9$  & $pp \rightarrow \ell N$ \\[1.5mm]

$\alpha_{LdQN}^{\ell qq}$ & $3.2$ & $2.2$ & $pp \rightarrow \ell N$   \\[1.5mm]

$\alpha_{LNQd}^{\ell qq}$ & $1.6$  & $3.2$  & $pp \rightarrow \ell N$   \\[1.5mm]

$\alpha_{QuNL}^{qq \ell}$ & $0.67$ & $4.9$ & $pp \rightarrow \ell N$ \\
\bottomrule
\end{tabular}
\caption{Summary of limits on four-fermion operators from LHC processes and observables they result from. Note that $\Lambda = 4 \tev$ is assumed for the limits
on $\alpha$ to stay within the range of validity of the EFT. 
%\vspace{0.5cm}
}
\label{tab:lim_LHC}
\end{table}
\begin{table}[tbh]
\centering
\begin{tabular}{lccc}
\toprule
 coefficient & $\alpha_\text{max}$ for $\Lambda= 1 \tev$ & $\Lambda_\text{min}$  [TeV] for $\alpha=1$ & observable\\
 \midrule
$\alpha_{eN}^{\ell \ell}$ & $0.93$ & $1.04$ & $ee \rightarrow N N$\\[1.5mm]
$\alpha_{LN}^{\ell \ell}$ & $0.93$ & $1.0$  & $ee \rightarrow N N$  \\[1.5mm]

%$\alpha_{QN}^{qq}$ & $0.048$ & $4.6$ & $pp \rightarrow NN$  \\[1.5mm]

%$\alpha_{uN}^{qq}$ & $0.058$ & $4.2$ & $pp \rightarrow NN$  \\[1.5mm]

%$\alpha_{dN}^{qq}$ & $0.078$ & $3.6$ & $pp \rightarrow NN$ \\[1.5mm]
$\alpha_{duNe}^{qq\ell}$ & $7.7 \times 10^{-4}$ & $36$ & $\pi \rightarrow \ell N$ \\
$\alpha_{duNe}^{qq\tau}$ & $0.49$ & $1.4$ & $\tau \to  \pi N$ \\
\bottomrule
\end{tabular}
\caption{Summary of limits on RRRR and LLRR operators and observables they result from, assuming $m_N = 0.1 \gev$.%\vspace{0.5cm}
}
\label{tab:lim_RRRR}
\end{table}
\begin{table}[tbh]
\centering
\begin{tabular}{lccc}
\toprule
 coefficient & $\alpha_\text{max}$ for $\Lambda= 1 \tev$ & $\Lambda_\text{min}$  [TeV] for $\alpha=1$ & observable\\
 \midrule
$\alpha_{LNLe}^{\ell \ell \tau}, \, \alpha_{LNLe}^{\tau \tau \ell}$ & $1.5$ & $0.82$ & $\tau \rightarrow \ell N \nu$ \\
$\alpha_{LNLe}^{\ell \tau \tau} , \, \alpha_{LNLe}^{\tau \tau \ell}$ & $1.7$ & $0.77$ & $\tau \rightarrow \ell N \nu$ \\
$\alpha_{LNLe}^{\ell \tau \ell} , \, \alpha_{LNLe}^{\tau \ell \ell}$ & $0.60$ & $1.3$ & $ee \rightarrow N \nu$ \\
$\alpha_{LNLe}^{\ell \ell \ell}$ & $0.52$ & $1.4$ & $ee \rightarrow N \nu$ \\[1.5mm]

%$\alpha_{LdQN}^{\ell qq}$ & $0.20$ & $2.2$ & $pp \rightarrow \ell N$   \\[1.5mm]

$\alpha_{LNQd}^{\ell qq}$ & $0.042$ & $4.9$ & $\pi \rightarrow \ell N \nu$  \\[1.5mm]

%$\alpha_{QuNL}^{qq \ell}$ & $0.042$ & $4.9$ & $pp \rightarrow \ell N$ \\
$( \alpha_{QuNL}^{qq\tau} -  \alpha_{LNQd}^{\tau qq} )$ & $0.30$ & $1.8$ & $\tau \to \pi N$ \\
\bottomrule
\end{tabular}
\caption{Summary of limits on LRRL operators and observables they result from, assuming $m_N = 0.1 \gev$. }
\label{tab:lim_LRRL}
\end{table}

Overall, many of the Wilson coefficients of the $\nu$SMEFT parameter space can already 
be constrained to $\alpha/\Lambda^2 \lesssim 1/\tev^2 $.
Our bounds are comparable to those obtained for very light RH neutrinos effectively stable at detector scales, from both LHC searches~\cite{Alcaide:2019pnf} as well as beta decay experiments~\cite{Bischer:2019ttk}. (Note however that this latter reference uses a slightly different operator basis, so the comparison is not inmediate.)
We should be aware, however, that some of these constraints are only valid 
for relatively small RH neutrino masses, \textit{e.g.}~$m_N < m_\tau$ or even $m_N < m_\pi$.
Moreover, we note that out of the 37 independent coefficients in our $\nu$SMEFT 
four-fermion Lagrangian, 
$17$ are still entirely unconstrained 
after our analyses in section~\ref{sec:searches}, namely 
\begin{equation}
\begin{split}
\alpha_{NN} , \,  &\alpha_{eN}^{\ell \tau}   , \, \alpha_{LN}^{\ell \tau} , \,
\alpha_{eN}^{\tau \tau} , \, \alpha_{LN}^{\tau \tau} , \, \alpha_{LNLe}^{\tau \tau \tau} , \, \alpha_{LdQN}^{\tau qq}  
\\ 
 &\alpha_{duNe}^{bt\tau } , \, \alpha_{LNQd}^{\tau 3b}  , \, \alpha_{LdQN}^{\tau b3} , \, \alpha_{LNQd}^{ 3b\tau}  
 \\
 \alpha_{uN}^{tt}   , \, &\alpha_{QN}^{33} , \,
 \alpha_{duNe}^{bt\ell } , \, \alpha_{LNQd}^{\ell 3b}  , \, 
 \alpha_{LdQN}^{\ell b3}  , \, \alpha_{QuNL}^{3t\ell}  \, .
\end{split}
\label{eq:unconstrained}
\end{equation}
While some of these operator coefficients, 
for instance those involving only the RH neutrinos
and $\tau$ leptons, will be difficult to constrain, dedicated searches 
will be able to probe further directions of our parameter space. 
In the next sections, we will point out further possibilities to probe NP triggered by some
of the coefficients in Eq.~\eqref{eq:unconstrained} using rare tau and top
decays. 

%---------------------------------------------------------------------------

\section{Projections for rare tau decays }
\label{sec:proj_tau}

The operators $\ope_{eN}$ and $\ope_{lN}$ contribute to the $\tau$ decay width in the 
$\tau \rightarrow \ell N N \rightarrow \ell \gamma \gamma \nu \nu$ channel:
\begin{equation}
\Gamma (\tau \rightarrow \ell N N \rightarrow \ell \gamma \gamma \nu \nu)|_{m_N=0} =
		\frac{m_\tau^5}{512 \pi^3 \Lambda^4} \left[  (\alpha_{eN}^{\ell \tau})^2 + (\alpha_{lN}^{\ell\tau})^2   \right] \, .
\end{equation}
The decay width includes a factor $3$ to account for the RH neutrino flavours.
The mass dependence of this decay channel is given by 
\footnote{We display the $m_N$ dependence of Eq.~\eqref{eq:width_tau_mN_NN} 
 (scaled by a factor $1/4$) in 
Appendix~\ref{app:mass_dep}, in the left panel of Fig.~\ref{fig:tauDecay_mN}.}
\begin{equation}
\left. \frac{\Gamma (x = \frac{m_N}{m_\tau})}{\Gamma(m_N=0)} \right|_{\tau \rightarrow \ell N N}
 = \sqrt{1-4x^2} (1 -14 x^2 - 2 x^4 - 12 x^6) + 48 x^4 (1-x^4) 
 \arccoth \left( \frac{1}{\sqrt{1-4x^2}} \right) \, .
\label{eq:width_tau_mN_NN}
\end{equation}

To the best of our knowledge, there are no experimental bounds on 
$\tau  \rightarrow \ell \gamma \gamma \nu (\nu)$.
In the SM, the contribution to this channel comes from $\tau \rightarrow \ell \nu \nu \gamma \gamma$,
\textit{i.e.}~two extra photons radiated in the decay $\tau \rightarrow \ell 
\nu \nu$.
We expect the main backgrounds to this channel to come from mistags and fakes, 
compare Ref.~\cite{Miyazaki:2007jp}, and will 
leave a dedicated study of this signature to experimentalists. 
To estimate the experimental sensitivity for this channel we can compare the uncertainties on the 
branching ratio of other $\tau$~decay channels in Ref.~\cite{Tanabashi:2018oca}, 
see also Tab.~\ref{tab:exp_tauDecays}.
We find that the uncertainties on $\text{BR}(\tau \rightarrow e \nu \nu)$ and 
$\text{BR}(\tau \rightarrow e \gamma \nu \nu)$ are $\sigma_\text{BR}=0.04\%$ 
and $\sigma_\text{BR}=0.05\%$, respectively. 
For decays to a muon, $\text{BR}(\tau \rightarrow \mu \nu \nu)$ with or without an extra photon, as well as for decays to a $\pi^0$ with 
subsequent decays to photons, the uncertainty on the branching ratio is (well) below $4\times10^{-4}$. 
Therefore, we will conservatively assume an absolute experimental uncertainty 
of $\sigma_\text{BR}= 0.05\%$ on the channel $\tau \rightarrow \ell \gamma \gamma \nu (\nu)$ 
which translates into a $\pm 1.1 \times 10^{-15} \gev$ uncertainty on the experimental decay width. 

For the limit setting, we allow the BSM contribution to the decay width to reach twice the 
assumed experimental uncertainty, \textit{i.e.}~$\Delta \Gamma^\text{BSM} = 2.3 
\times 10^{-15} \gev$.
For $m_N=0.1\gev$ and $\Lambda=1\tev$, the Wilson coefficients $\alpha_{eN}^{\ell \tau}$ and 
$\alpha_{LN}^{\ell \tau}$ can be constrained to $|\alpha_{eN}^{\ell \tau}|, \, |\alpha_{L N}^{\ell \tau}| <~1.5$.
If the experimental uncertainty on the branching ratio can be reduced to $10^{-5}$, the resulting limit 
is $|\alpha_{eN}^{\ell \tau}|, \, |\alpha_{LN}^{\ell \tau}| < 0.21$.
The mass dependence of these limits is shown in the right panel of Fig.~\ref{fig:tauDecay_limits_proj}.

\begin{figure}
	\centering
	\includegraphics[width=.45\textwidth]{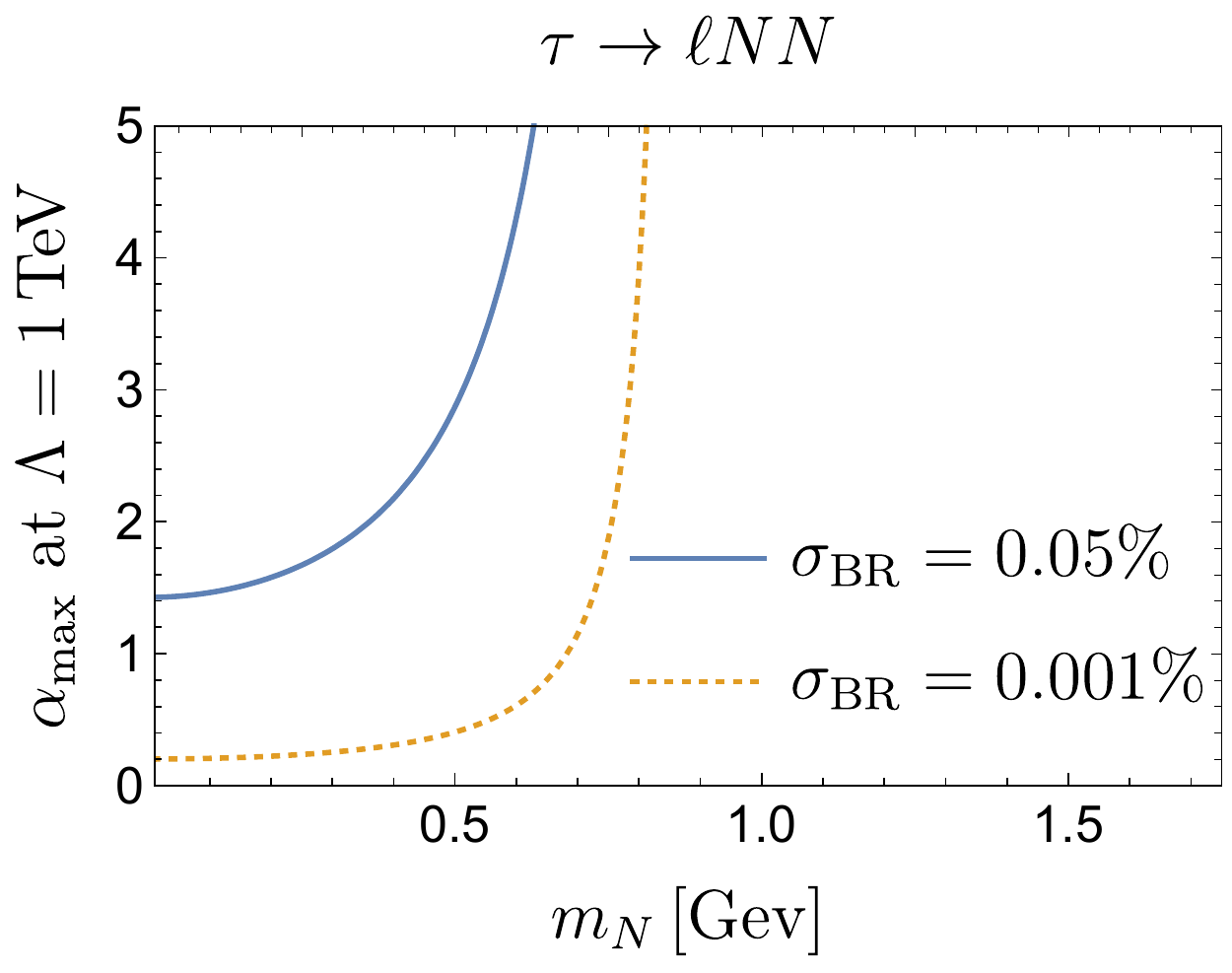}
	\caption{RH neutrino mass dependence of the projected limits 
	on $\alpha_{eN}^{\ell\tau}$ and $\alpha_{LN}^{\ell\tau}$
	from $\tau$~decays in the $\tau \rightarrow \ell N N$ channel. 
	We show the limits for two different assumptions on the experimental 
uncertainty of 
	the branching ratio.
	The mass on the charged leptons has been neglected.}
	\label{fig:tauDecay_limits_proj}
\end{figure}

\section{Projections for rare top decays}
\label{sec:top}
The weak sensitivity of current analyses to operators involving the top quark 
(see the end of section~\ref{sec:lepton_photon_etmiss})
suggests that dedicated searches for signals triggered by these operators must 
be developed. We propose one such search strategy in top pair production, 
with one of the top quarks decaying as $t\to b\ell N, N\to\gamma\nu$, and the 
other via the dominant SM channel, $t\to bW$.
We focus on the signal ensuing from the hadronic decay of the $W$.
%
%\ab{[Do you think we should repeat a cross section (or number of event) 
%parametrization here to be consistent with the other searches? Just to be 
%reminded of the interfering operators...]}

The background is dominated by the
process $t\overline{t}\gamma$. 
For event simulation, we employ the same tool chain as above, 
compare section~\ref{sec:searches}. 
We simulate the corresponding samples at $\sqrt{s}=13$ TeV 
with no parton level cuts
for the signal and enforcing $p_T^\gamma > 10$ GeV for the background.
The tree-level cross section of the signal, up to the rare top branching ratio, 
is
$\sigma_s \approx 240$ pb for a top mass $m_t = 172.5$ GeV. For the background 
we obtain $\sigma_b \approx 0.68$ pb. We rescale both cross sections by an 
approximated NLO 
$\alpha_s$ K-factor
of 1.5~\cite{Alwall:2014hca} and we 
neglect detector effects.
We implement the following search strategy: First, we require events to have 
exactly one
(light) lepton with $p_T^\ell > 25$ GeV and $|\eta_\ell| < 2.5$, exactly one 
isolated
photon with $p_T>12$ GeV and at least three jets with $p_T > 30$ GeV, of which 
exactly
two must be $b$-tagged.
\footnote{
A photon is isolated if the sum of the transverse momentum of all leptons and 
hadrons in
a cone
of $\Delta R < 0.3$ around the photon candidate is smaller than 10\% of its 
transverse
momentum.
Jets are clustered using the anti-k$_t$ algorithm~\cite{Cacciari:2008gp} with 
$R=0.4$. All 
hadrons
and photons which are either not isolated or have a low transverse momentum 
$p_T^\gamma< 12$ GeV are 
considered in the 
clustering process (leptons are not). 
We assume a jet to be a $b$-jet candidate if there is a $B$-meson within a cone 
of
$\Delta R = 0.5$ of its four-momentum. The $b$-tagging efficiency is set to 
$0.7$.}
In addition, we require $\etmiss > 30$ GeV. 
We will refer to this set of restrictions as \textit{basic cuts}. 

\begin{figure}[t]
\includegraphics[width=0.49\columnwidth]{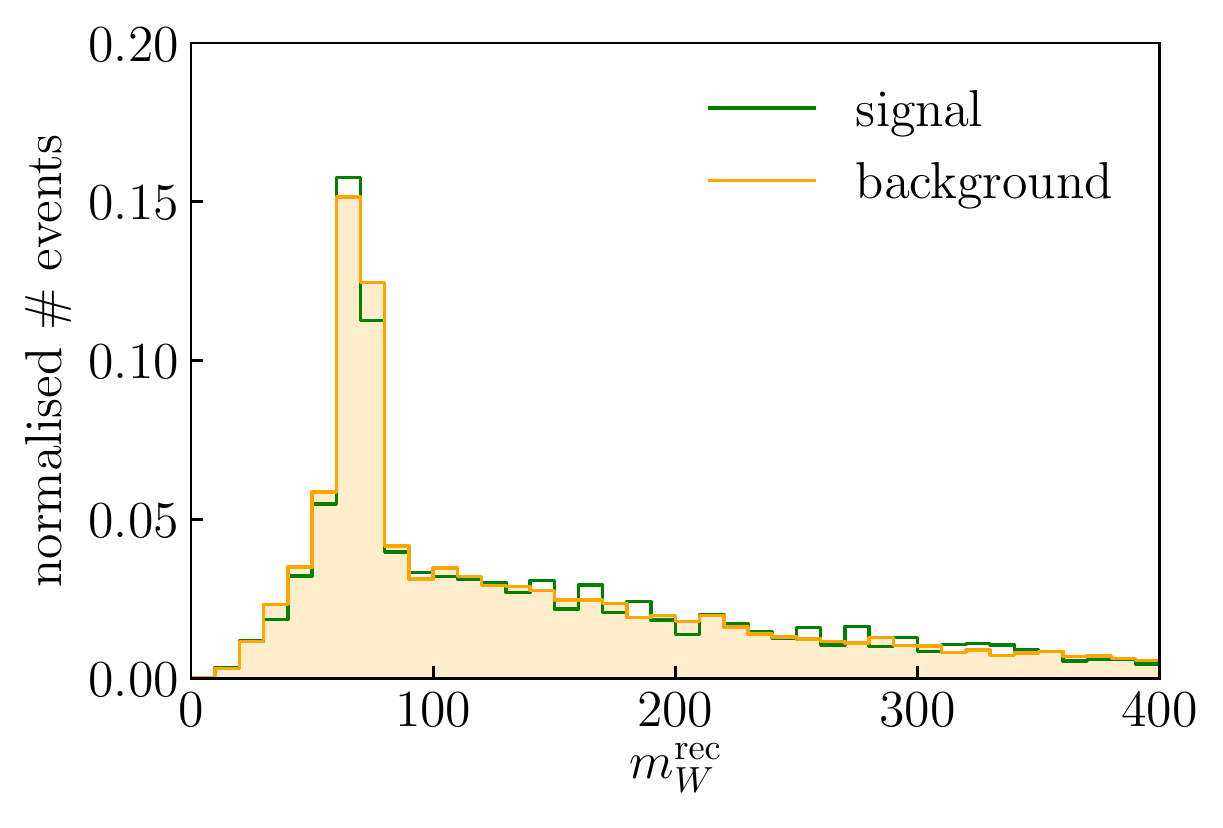}
\includegraphics[width=0.49\columnwidth]{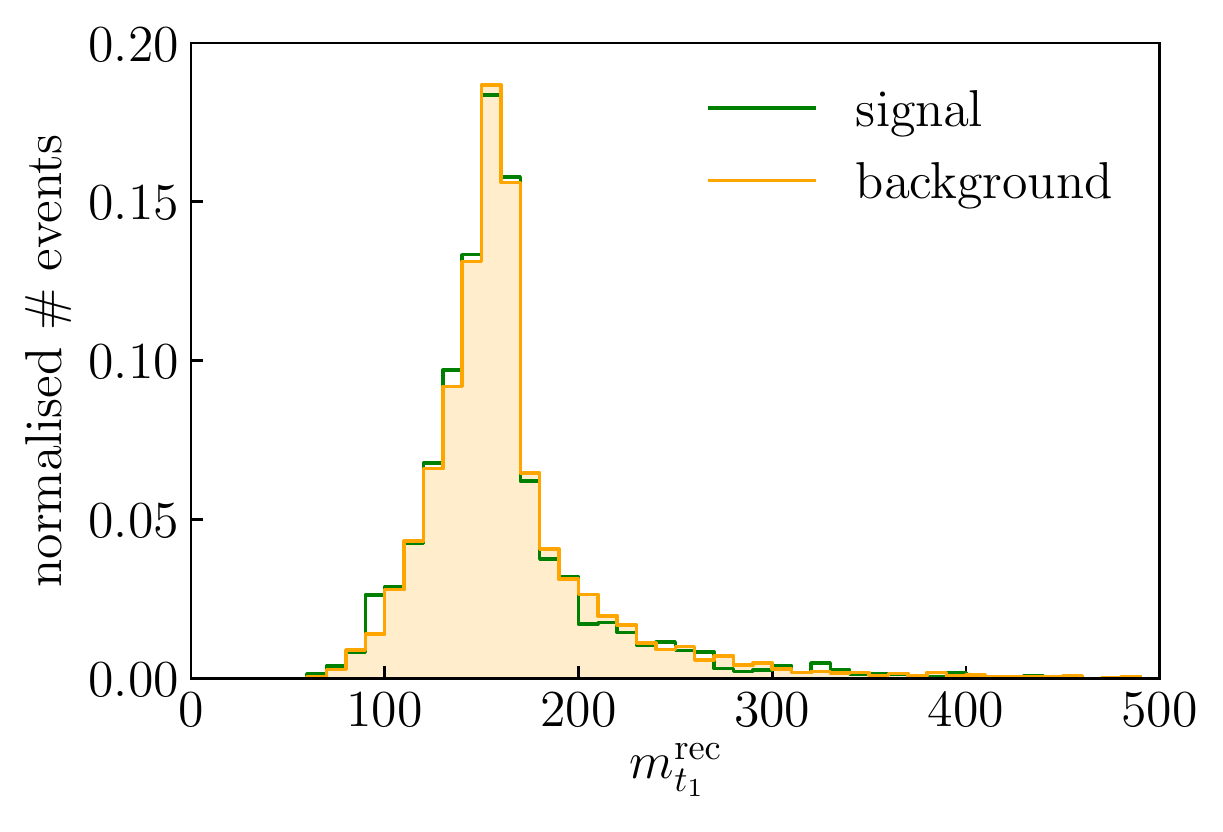}
\includegraphics[width=0.495\columnwidth]{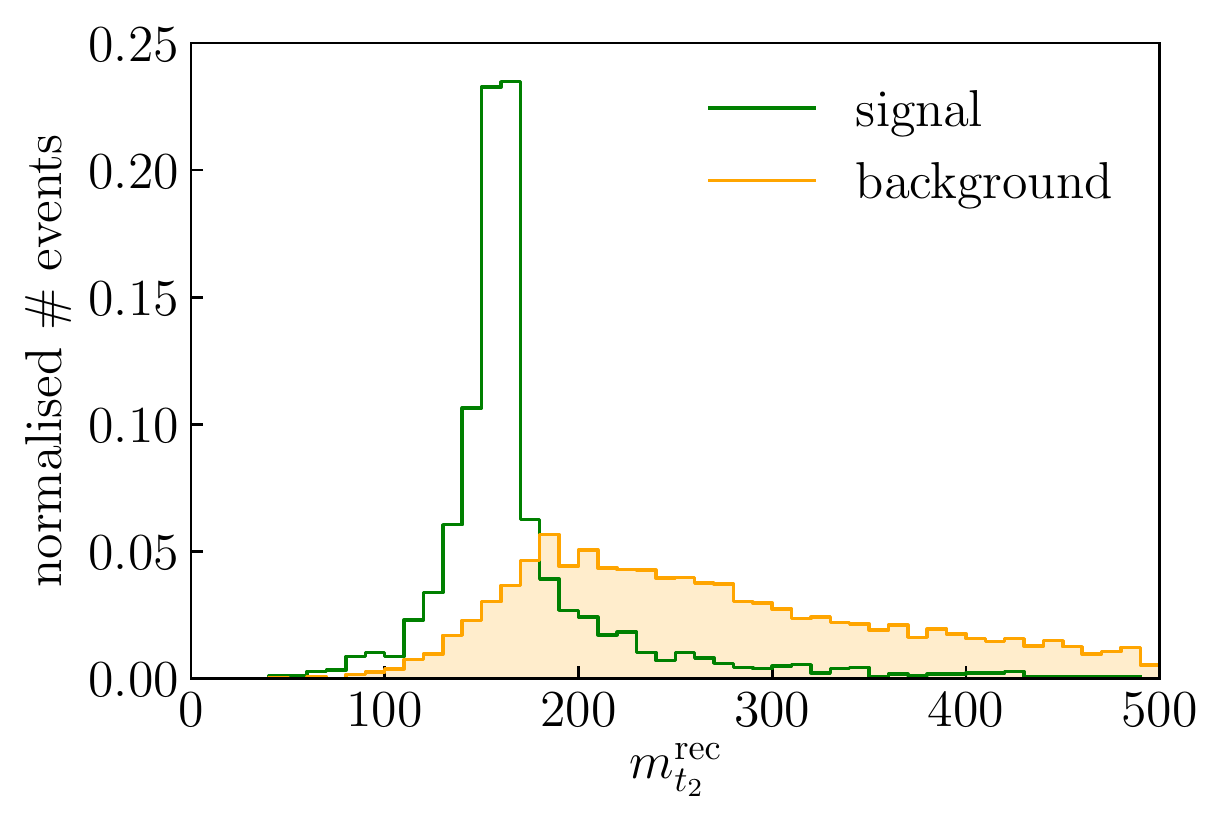}
\includegraphics[width=0.49\columnwidth]{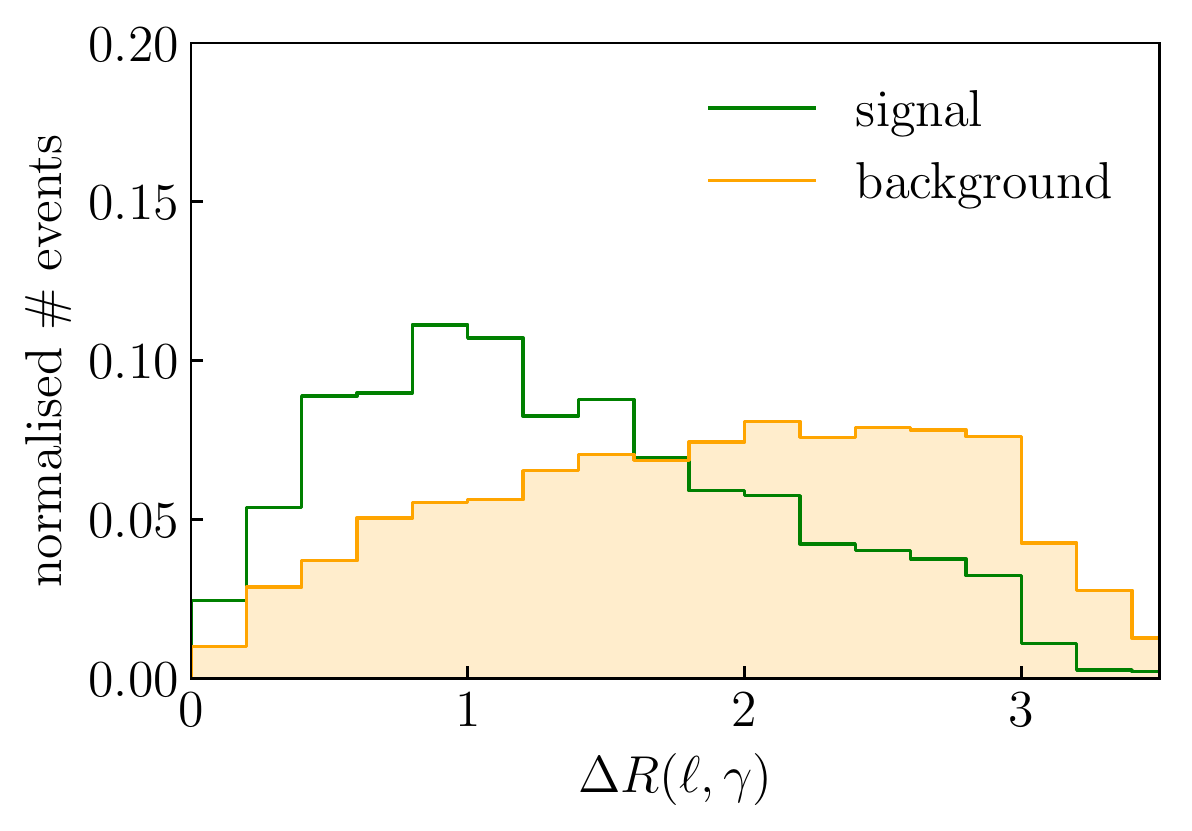}
\caption{Normalised distribution of different observables in $t \bar{t}$ production. Top left: reconstructed $W$ mass after the basic cuts. Top right: reconstructed mass of the SM-decaying top after the cut on $m_W^\text{rec}$. Bottom 
left: reconstructed mass of the rare decaying top after the 
cut on $m_{t_1}^\text{rec}$. Bottom right: 
angular separation between the lepton and the photon after the cut on 
$m_{t_1}^\text{rec}$. In all cases, the signal (background) appears in green 
(orange).
}
\label{fig:mWrec}
\end{figure}

In a second step, we reconstruct the $W$ boson from the two leading light jets. 
The
normalised distribution of its invariant mass $m_W^\text{rec}$ is shown in the 
upper
left panel of Fig.~\ref{fig:mWrec}
in both the signal and the background. We require $m_W^\text{rec}$ to lie
in the window $m_W^\text{rec} \in [50, 120]$ GeV.

We subsequently reconstruct the SM top from the $W$ and the $b$-tagged jet 
closer to it in $\Delta R$. 
The normalised
distribution of the corresponding mass $m_{t_1}^\text{rec}$ in both the signal 
and
the background is depicted in the upper right panel of the 
Fig.~\ref{fig:mWrec}. 
We require this observable to lie in the window $[100, 200]$ GeV. 

Finally, we reconstruct two variables that can discriminate well signal from 
background.
The first one is the invariant mass of the reconstructed leptonic top, 
$m_{t_2}^\text{rec}$.
This top is built from the lepton, the remaining $b$-tagged jet, the photon and 
the
neutrino. (The $x$ and $y$ components of the neutrino are identified with the
respective components of the missing energy; the longitudinal component is 
obtained
under the \textit{collinear assumption} by which the neutrino and the photon 
three-momenta
are aligned because they are the two decay products of a very light particle, 
$N$.)
This observable peaks around the top quark mass $\sim 172$ GeV in the signal 
while it is
more spread in the background; see the bottom left panel of 
Fig.~\ref{fig:mWrec}.

The second discriminating variable is the $\Delta R$ separation between the 
lepton and
the photon, $\Delta R(\ell,\gamma)$. Because these two objects originate from the 
decay of the same top quark in the
signal, this variable is peaked to smaller values in the signal than in the 
background,
where it is flatter; see the bottom right
panel of the aforementioned Fig.~\ref{fig:mWrec}. 

These two variables are however highly correlated. Thus, for example, a cut on 
$m_{t_2}^\text{rec} < 200$ GeV
reduces significantly the difference between signal and background in 
$\Delta(\ell,\gamma)$.
For this reason, we propose two different statistical analyses, each using just 
one  of 
these variables
at a time. First, we just count the number of events passing the cut on 
$150\,\, 
\text{GeV}< m_{t_2}^\text{rec}<200$ GeV.
The efficiencies for selecting signal and background events in this region are
$\sim 0.013$ and $\sim 0.0073$, respectively. (The small difference between 
signal and background is mostly due to
the different parton-level cuts.)
Thus, for a luminosity of ${\cal L}=3$ ab$^{-1}$  and assuming a 10\% 
uncertainty
on the background, we obtain that $\mathcal{B}(t\to b\ell N) > 1.6\times 
10^{-4}$ can be probed at the 95\% CL
upon using the CL$_s$ method. 

A potentially more robust analysis relies on the asymmetry
\begin{equation}
A = \frac{N_+ - N_-}{N_+ + N_-} = \frac{N(\Delta R(\ell, \gamma) > 2) - 
N(\Delta 
R(\ell, \gamma) < 2)}{N(\Delta R(\ell, \gamma) > 2) + N(\Delta R(\ell, \gamma) > 
2)}\,.\label{eq:ass}
\end{equation}
Systematic uncertainties are expected to cancel in this ratio. The efficiency 
for selecting events in the
region $N_+ (N_-)$ (defined as the ratio of events that pass all cuts in each
region over the total number of events before the basic cuts) is of about $0.0055$ ($0.014$) in the signal and $	
0.028$ ($0.021$) in the 
background. 

In the left panel of Fig.~\ref{fig:topResults} we show the CL (in number of 
standard deviations) to which the signal can be probed depending on 
$\mathcal{B}(t\to b\ell N)$ and for two different assumptions on the collected 
luminosity. In the right panel, we plot the luminosity required to test the 
signal at two different levels of confidence, again as a function of the top's 
rare decay branching ratio.

For ${\cal L} = 3$ ab$^{-1}$, the value of $A$ in the signal departs by more than two 
sigmas from the 
SM, \textit{i.e.} $A_s < A_b - 2\sigma(A_b)$, for
$\mathcal{B}(t\to b\ell N) > 6.6\times 10^{-5}$.
\begin{figure}[t]
\raisebox{0.9mm}[0pt][0pt]{\includegraphics[width=0.49\columnwidth]
{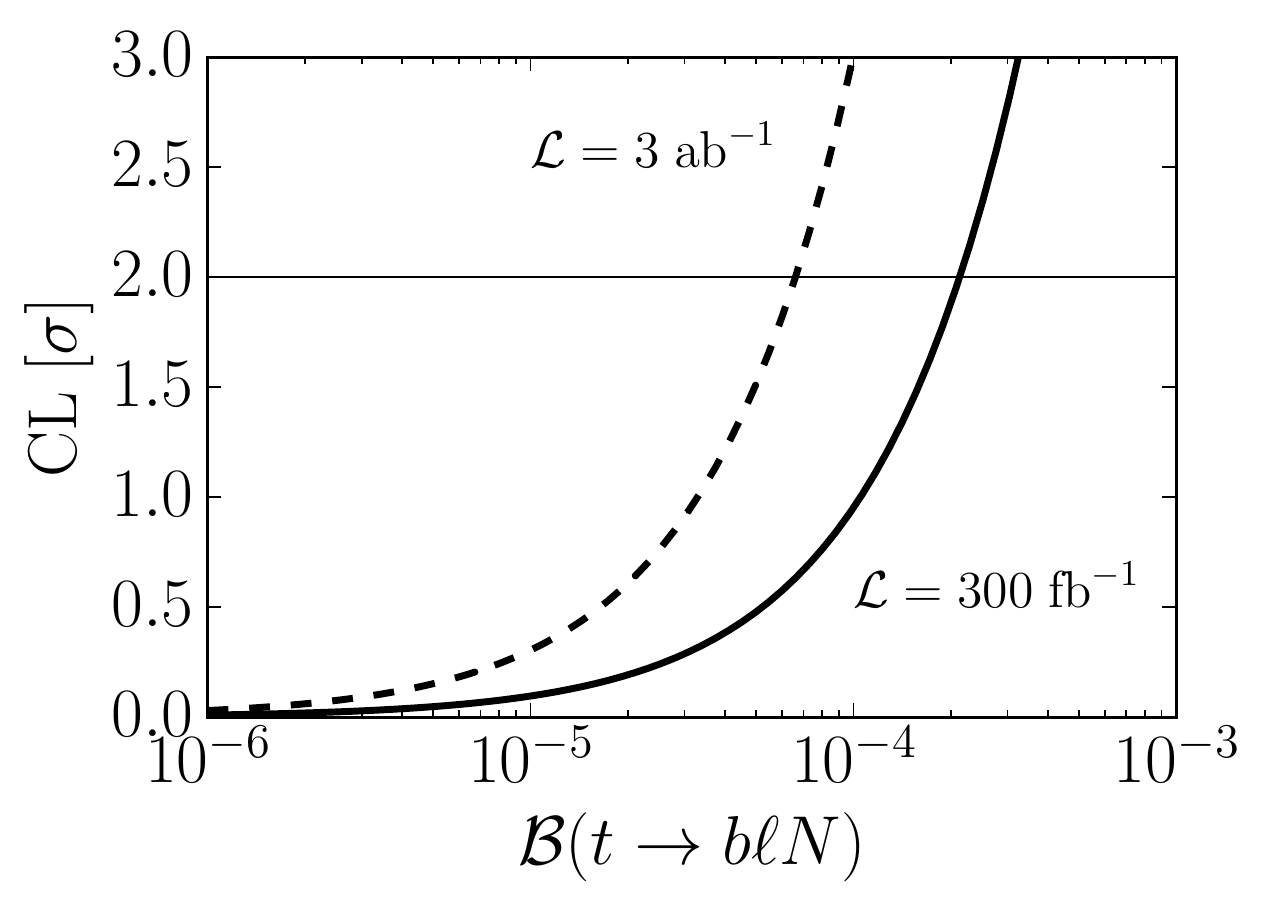}}
\includegraphics[width=0.49\columnwidth]{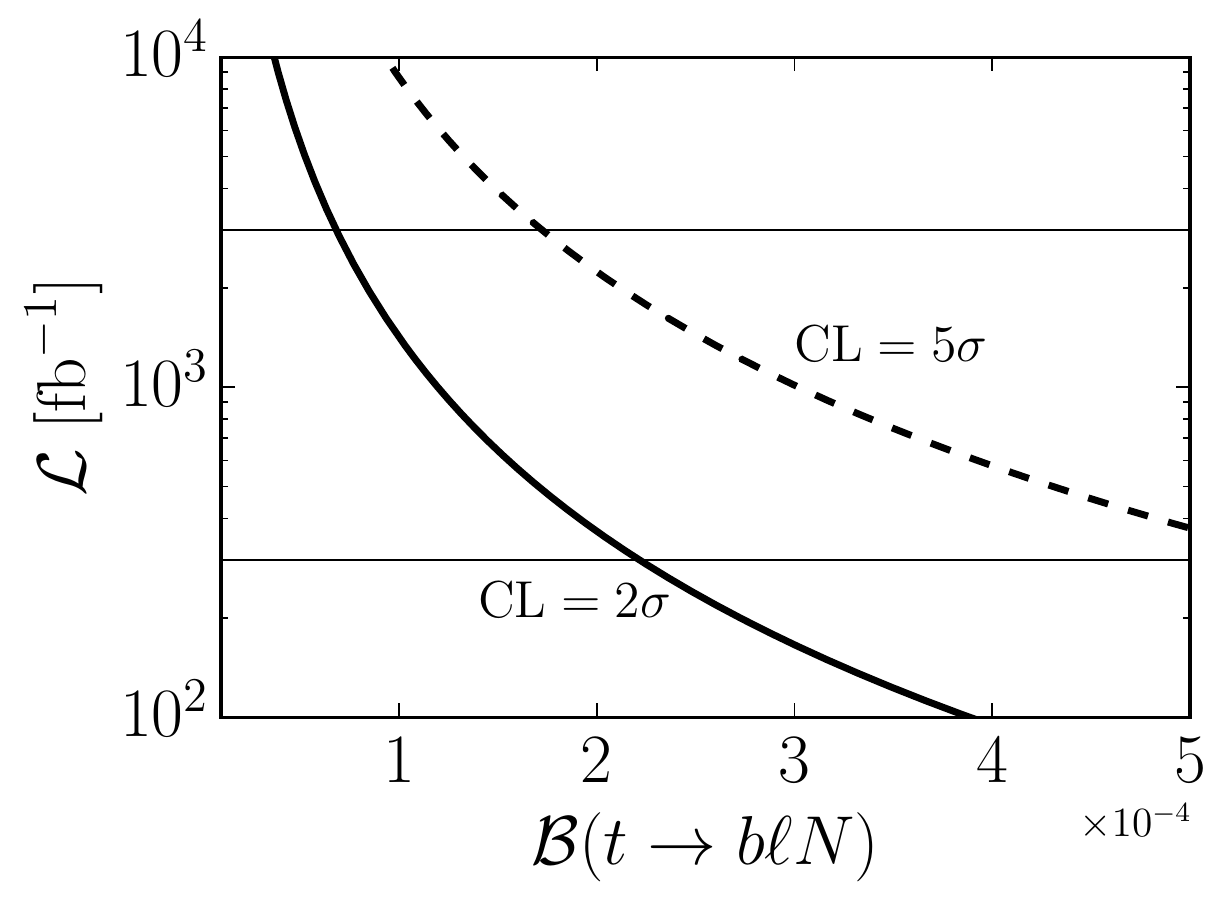}
\caption{Left: LHC sensitivity to $t\to b\ell N$ as a function of the 
branching ratio for two values of the collected 
luminosity. Right: Luminosity required to probe $t\to 
b\ell N$ to $2\sigma$ and $5\sigma$ as a function of the 
branching ratio. In both cases we rely on the analysis based
on the asymmetry defined in Eq.~\eqref{eq:ass}.
}
\label{fig:topResults}
\end{figure}
Under the flavour-universality assumption (the top decays into both $eN_i$ and 
$\mu N_i$, with $i=1,2,3$), and using Eq. (2.27) in 
Ref.~\cite{Alcaide:2019pnf}, the 
expected limit on $\mathcal{B}(t\to b\ell N)$ translates into 
$|\alpha_{duNe}^{bt\ell}|<2.3$, $|\alpha_{QuNL}^{3t\ell}|<4.5$ and 
$|\alpha_{LdQN}^{\ell b3}|$, 
$|\alpha_{LNQd}^{\ell 3b}|<5.1$, for $\Lambda=1$ 
TeV. For setting bounds on the last two operators we have 
marginalised over the interfering one.

\section{Conclusions}
\label{sec:conclusions}
In this paper we have studied the phenomenology of the low-scale see-saw EFT, in
the regime in which the sterile neutrinos $N$ decay as $N\to\nu\gamma$.
With the aim of unravelling
in which directions of the parameter space new physics can hide, we have derived
constraints on the different Wilson coefficients, with special attention to
four-fermion operators as they can arise at tree level in UV completions of the
see-saw model.

For this goal we have relied on data from LHC searches for one lepton,
one photon and missing energy and two photons and missing energy; on 
measurements
of different pion and tau decays; as well as on LEP data from analyses of one or multiple
photons and missing energy. 
The strongest limits result from LHC searches and, in the low-$m_N$ regime, also from pion decays. Operator coefficients constrained from these 
processes obtain bounds of $\alpha/\Lambda^2 \lesssim 0.2 \tev^{-2}$. 
LEP limits are below $\alpha/\Lambda^2 \lesssim 1 \tev^{-2}$.

We note that, in deriving these bounds, we have assumed flavour universality in
$N$ as well as in the light fermions and quarks; and we allowed LFV only in 
tau-to-light-lepton
transitions. 
Nonetheless, our results can trivially be interpreted under different
assumptions. For example, if the three $N$ flavours couple differently to 
the SM 
fermions,
then the bound on $\alpha_{uN}^{1111}$ is just $\sqrt{3} \approx 1.73$ times weaker than the one
we provide on $\alpha_{uN}^{qq}$. Likewise, if moreover flavour-universality in
the light
quarks is abandoned, the bound on $\alpha_{uN}^{2211}$ can be estimated from $c\bar{c} \to NN$ versus
$u\bar{u}\to NN$ as
$4.6$ times the limit on $\alpha_{uN}^{qq}$ due to the PDF suppression. 
 
Applying our results to UV models where several operators arise simultaneously
(and therefore the bounds are strengthened) is also straightforward, as we have provided
master equations to straightforwardly predict the number of signal events in the
different signal regions as well as quoted the upper limit on the latter in each
case.

Still, there are operators coefficients that current data do not bound. These include 
the parameters $\alpha_{eN}^{\ell \tau}$ and $\alpha_{LN}^{\ell \tau}$
which trigger the tau decay $\tau\to \ell\gamma\gamma\nu\nu$. 
The resulting limits very much depend on the estimated experimental 
sensitivity of the branching ratio, which we conservatively assume to 
be $\sim 0.05\%$. The emerging bounds are $|\alpha_{eN}^{\ell \tau}|, \, |\alpha_{L N}^{\ell \tau}| <~1.5$ for $\Lambda = 1 \tev$. To push these limits below 
$\alpha/\Lambda^2 \lesssim 1  \tev^{-2}$, an experimental sensitivity on the 
branching ratio below $\sigma_\text{BR} \lesssim 0.023 \%$ has to be reached.
Other operator coefficients that are very weakly constrained by current data are 
$\alpha_{duNe}^{bt\ell}$, $\alpha_{LdQN}^{\ell b 3}$, 
$\alpha_{LNQd}^{\ell 3 b}$ and $\alpha_{QuNL}^{3t\ell}$, 
which
drive the top decay $t\to b\ell\gamma\nu$. We have provided a dedicated analysis
to test this channel in top pair production at the LHC, and found that branching
ratios as small as $6.6\times 10^{-5}$ could be probed at the 95\% CL in the high-luminosity phase.
This in turn translates to a potential upper bound on $\alpha_{duNe}^{bt\ell}$ 
of $\sim 2.3$ for
$\Lambda = 1$ TeV; and about twice weaker for the others.

In total, $11$ out of $37$ four-fermion operator coefficients in our 
$\nu$SMEFT Lagrangian 
remain unconstrained even after our additional analyses. 
In particular, this concerns operator coefficients describing couplings of tau leptons 
to the third quark generation, which could potentially be bounded by analyses 
of top decays to tau leptons, photons and missing
energy~\footnote{The operator $\mathcal{O}_{HNe}$ might be also tested in 
top decays, following a strategy similar to that in 
Ref.~\cite{Liu:2019qfa}.}. Coefficients describing $\tau\tau NN$, $\tau \tau 
tt$ and $\tau \tau bb$ couplings
are not constrained either.
We leave studies to bound these directions of the parameter space for future work.

Altogether, our work highlights in particular the importance of performing dedicated searches
for new rare tau and top decays.

\section*{Acknowledgements}
AB and MS acknowledge support by the UK Science and Technology Facilities 
Council (STFC) under grant  ST/P001246/1. MC is supported by the Spanish 
MINECO under the Juan de la Cierva programme as 
well as by the Ministry of Science
and Innovation under grant number FPA2016-78220-C3-3-P, 
and by the Junta de Andaluc{\'\i}a grants FQM 101 and A-FQM-211-UGR18 (fondos 
FEDER). 

%-----------------------------------------------------------------------
\FloatBarrier
\newpage
\appendix
\section{Explicit Lagrangian}
\label{app:lag}

In order to further clarify our notation, we write here explicitly the full 
$\nu$SMEFT dimension-six Lagrangian indicating all independent Wilson 
coefficients according to our flavour assumptions.

The relevant bosonic Lagrangian is
\begin{align}\nonumber
 L &= \mc{\alpha_{HN}} \, \mathcal{O}_{HN}^{ii}%\\\nonumber
 +\mc{\alpha_{HNe}^{\ell}}\, 
(\mathcal{O}_{HNe}^{i1}+\mathcal{O}_{HNe}^{i2})%\\\nonumber
 +\mc{\alpha_{HNe}^{\tau}}\, \mathcal{O}_{HNe}^{i3}\\\nonumber
 &+\mc{\alpha_{NA}^{\ell}}\, 
(\mathcal{O}_{NA}^{1i}+\mathcal{O}_{NA}^{2i})%\\\nonumber
 +\mc{\alpha_{NA}^{\tau}}\, \mathcal{O}_{NA}^{3i}%\\\nonumber
 +\mc{\alpha_{NZ}^{\ell}}\, 
(\mathcal{O}_{NZ}^{1i}+\mathcal{O}_{NZ}^{2i})%\\\nonumber
 +\mc{\alpha_{NZ}^{\tau}}\, \mathcal{O}_{NZ}^{3i}\,,
\end{align}
with $i=1,2,3$. 

And for the relevant four-fermion operators we have (to be read in two columns):
\begin{equation*}
\begin{aligned}[tc]
 L &= \mc{\alpha_{NN}}\, \mathcal{O}_{NN}^{iijj} \\\nonumber
 & + \mc{\alpha_{eN}^{\ell\ell}}\, (\mathcal{O}_{eN}^{11ii}+\mathcal{O}_{eN}^{22ii}) \\\nonumber
 &+ \mc{\alpha_{eN}^{\ell\tau}}\, (\mathcal{O}_{eN}^{13ii}+\mathcal{O}_{eN}^{23ii}+\mathcal{O}_{eN}^{31ii}+\mathcal{O}_{eN}^{32ii}) \\\nonumber
 &+ \mc{\alpha_{eN}^{\tau\tau}}\, \mathcal{O}_{eN}^{33ii}\\\nonumber
 &+\mc{\alpha_{uN}^{qq}}\, (\mathcal{O}_{uN}^{11ii}+\mathcal{O}_{uN}^{22ii}) \\\nonumber
 &+  \mc{\alpha_{uN}^{tt}}\, \mathcal{O}_{uN}^{33ii} \\\nonumber
 &+ \mc{\alpha_{dN}^{qq}}\, (\mathcal{O}_{dN}^{11ii}+\mathcal{O}_{dN}^{22ii}) \\\nonumber
 &+  \mc{\alpha_{dN}^{bb}}\, \mathcal{O}_{dN}^{33ii}\\\nonumber
 &+\bigg[\mc{\alpha_{duNe}^{qq\ell}}\, (\mathcal{O}_{duNe}^{11i1} + \mathcal{O}_{duNe}^{11i2} + \mathcal{O}_{duNe}^{22i1} + \mathcal{O}_{duNe}^{22i2}) \\\nonumber
 &+ \mc{\alpha_{duNe}^{qq\tau}}\, (\mathcal{O}_{duNe}^{11i3} + \mathcal{O}_{duNe}^{22i3})\\\nonumber
 &+ \mc{\alpha_{duNe}^{bt\ell}}\, (\mathcal{O}_{duNe}^{33i1} + \mathcal{O}_{duNe}^{33i2}) \\\nonumber
 &+ \mc{\alpha_{duNe}^{bt\tau}}\, (\mathcal{O}_{duNe}^{33i3} + \mathcal{O}_{duNe}^{33i3}) + \text{h.c.}\bigg] \\\nonumber
 &+\mc{\alpha_{LN}^{\ell\ell}}\, (\mathcal{O}_{LN}^{11ii}+\mathcal{O}_{LN}^{22ii})\\\nonumber
 &+ \mc{\alpha_{LN}^{\ell\tau}}\, (\mathcal{O}_{LN}^{13ii}+\mathcal{O}_{LN}^{23ii}+\mathcal{O}_{LN}^{31ii}+\mathcal{O}_{LN}^{32ii})  \\\nonumber
 &+ \mc{\alpha_{LN}^{\tau\tau}}\, \mathcal{O}_{LN}^{33ii} \\\nonumber
 &+ \mc{\alpha_{QN}^{qq}}\, (\mathcal{O}_{QN}^{11ii}+\mathcal{O}_{uN}^{22ii})  \\\nonumber
 &+  \mc{\alpha_{QN}^{33}}\, \mathcal{O}_{QN}^{33ii}\\\nonumber
 \\\nonumber
 \\\nonumber
\end{aligned}
\qquad
\begin{aligned}[c]
 &+ \bigg[\mc{\alpha_{LNLe}^{\ell\ell\ell}}\, (\mathcal{O}_{LNLe}^{1i11} + \mathcal{O}_{LNLe}^{2i22})\\\nonumber
 &+ \mc{\alpha_{LNLe}^{\ell\ell\tau}}\, (\mathcal{O}_{LNLe}^{1i13} + \mathcal{O}_{LNLe}^{2i23})\\\nonumber
 &+ \mc{\alpha_{LNLe}^{\ell\tau\ell}}\, (\mathcal{O}_{LNLe}^{1i31} + \mathcal{O}_{LNLe}^{2i32})\\\nonumber
 &+ \mc{\alpha_{LNLe}^{\ell\tau\tau}}\, (\mathcal{O}_{LNLe}^{1i33} + \mathcal{O}_{LNLe}^{2i33})\\\nonumber
 &+ \mc{\alpha_{LNLe}^{\tau\ell\ell}}\, (\mathcal{O}_{LNLe}^{3i11} + \mathcal{O}_{LNLe}^{3i22})\\\nonumber
 &+ \mc{\alpha_{LNLe}^{\tau\ell\tau}}\, (\mathcal{O}_{LNLe}^{3i13} + \mathcal{O}_{LNLe}^{3i23})\\\nonumber
 &+ \mc{\alpha_{LNLe}^{\tau\tau\ell}}\, (\mathcal{O}_{LNLe}^{3i31} + \mathcal{O}_{LNLe}^{3i32})\\\nonumber
 &+ \mc{\alpha_{LNLe}^{\tau\tau\tau}}\, \mathcal{O}_{LNLe}^{3i33}\\\nonumber
 &+ \mc{\alpha_{LNQd}^{\ell qq}}\, (\mathcal{O}_{LNQd}^{1i11} + \mathcal{O}_{LNQd}^{1i22} + \mathcal{O}_{LNQd}^{2i11} + \mathcal{O}_{LNQd}^{2i22})\\\nonumber
 & +\mc{\alpha_{LNQd}^{\ell 3b}}\, (\mathcal{O}_{LNQd}^{1i33} + \mathcal{O}_{LNQd}^{2i33})\\\nonumber
 & +\mc{\alpha_{LNQd}^{\tau qq}}\, (\mathcal{O}_{LNQd}^{3i11} + \mathcal{O}_{LNQd}^{3i22})\\\nonumber
 & +\mc{\alpha_{LNQd}^{\tau 3b}}\,  \mathcal{O}_{LNQd}^{3i33}\\\nonumber
 & +\mc{\alpha_{LdQN}^{\ell qq}}\, (\mathcal{O}_{LdQN}^{111i} + \mathcal{O}_{LdQN}^{122i} + \mathcal{O}_{LdQN}^{211i} + \mathcal{O}_{LdQN}^{22i})\\\nonumber
 &+\mc{\alpha_{LdQN}^{\ell b3}}\, (\mathcal{O}_{LdQN}^{133i} + \mathcal{O}_{LdQN}^{233i})\\\nonumber
 &+\mc{\alpha_{LdQN}^{\tau qq}}\, (\mathcal{O}_{LdQN}^{311i} + \mathcal{O}_{LdQN}^{322i})\\\nonumber
 &+\mc{\alpha_{LdQN}^{\tau b3}}\, \mathcal{O}_{LdQN}^{333i}\\\nonumber
 &+\mc{\alpha_{QuNL}^{qq\ell}}\, (\mathcal{O}_{QuNL}^{11i1} + \mathcal{O}_{QuNL}^{11i2} + \mathcal{O}_{QuNL}^{22i1} + \mathcal{O}_{QuNL}^{22i2})\\\nonumber
 &+\mc{\alpha_{QuNL}^{qq\tau}}\, (\mathcal{O}_{QuNL}^{11i3} + \mathcal{O}_{QuNL}^{22i3})\\\nonumber
 &+\mc{\alpha_{QuNL}^{3t\ell}}\, (\mathcal{O}_{QuNL}^{33i1} + \mathcal{O}_{QuNL}^{33i2})\\
& +\mc{\alpha_{QuNL}^{3t\tau}}\, \mathcal{O}_{QuNL}^{33ie} + \text{h.c.}\bigg]~.
\end{aligned}
\end{equation*}

\section{Mass dependence of pion and tau decay widths}
\label{app:mass_dep}
In Fig.~\ref{fig:pionDecay_mN}, we explicitly show the mass dependence 
of the pion decay width in the $\pi \to \ell N$ channel for operators with 
vector and pseudo-scalar couplings. 
\begin{figure}[h!]
	\centering
	\includegraphics[width=.45\textwidth]{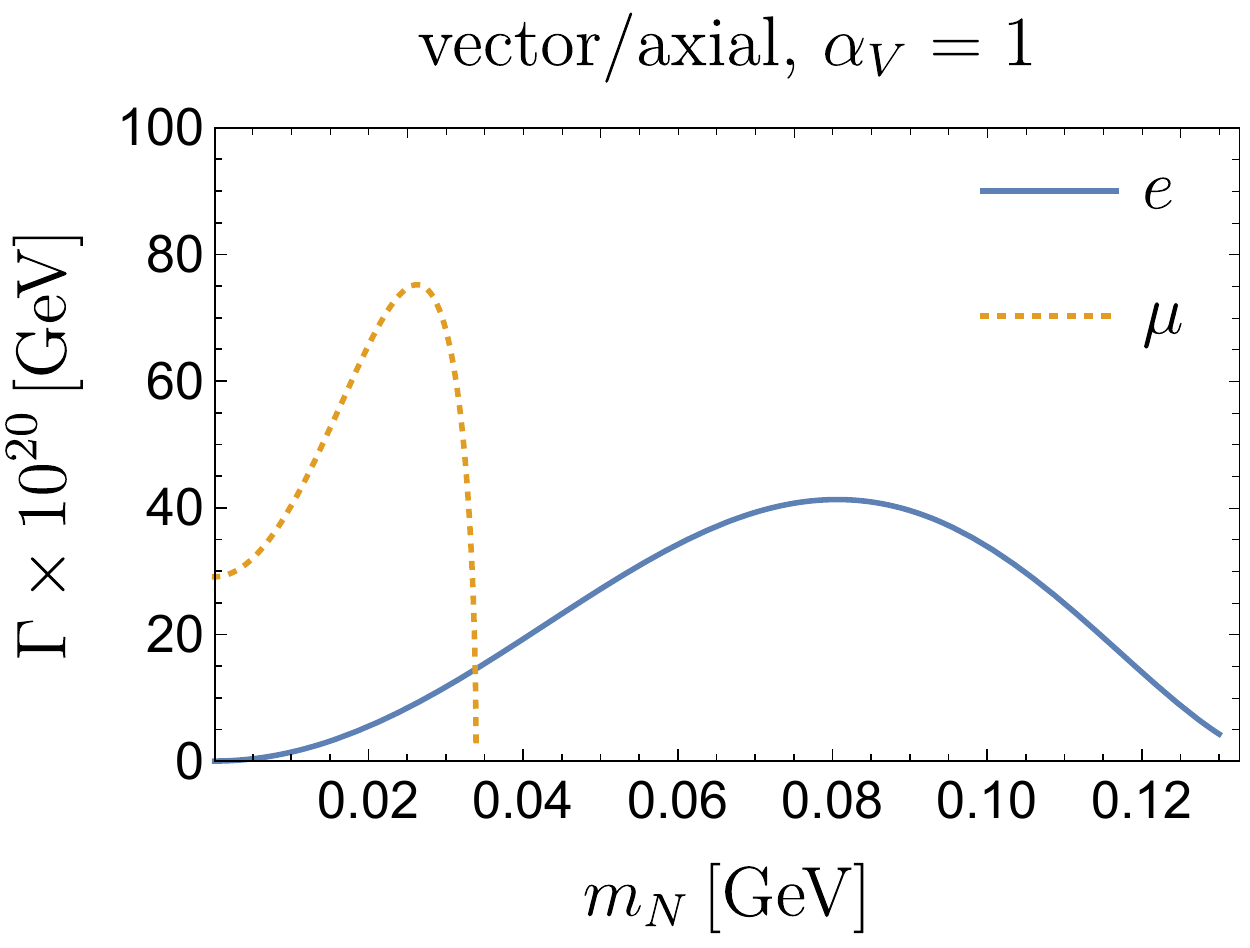}
	\quad
	\includegraphics[width=.45\textwidth]{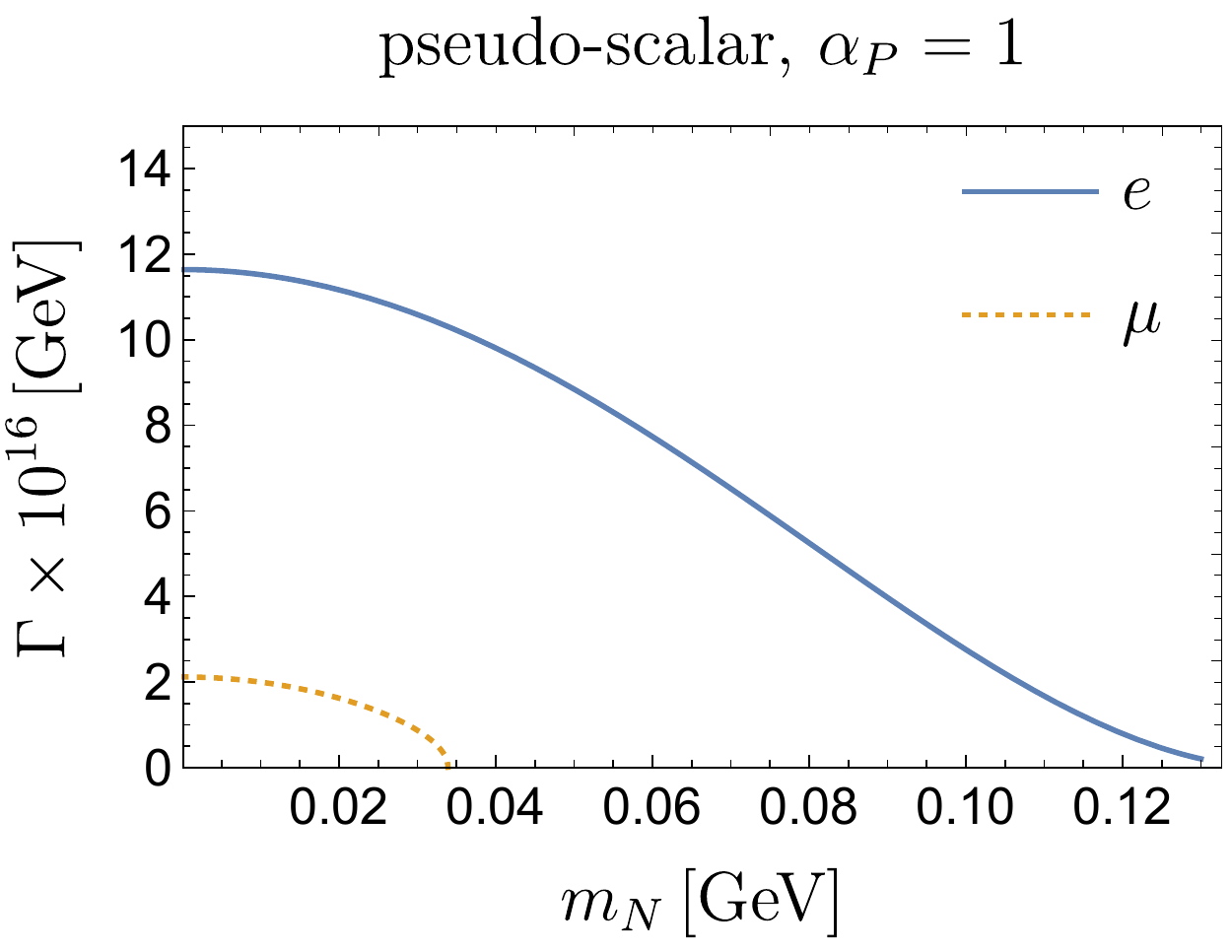}
	\caption{Dependence of the pion decay width on the neutrino mass $m_N$ for
	operators with axial (left) and pseudo-scalar (right) couplings.}
	\label{fig:pionDecay_mN}
\end{figure}

In Fig.~\ref{fig:tauDecay_mN}, we explicitly show the mass dependence 
of the $\tau$~decay width in the $\tau \to \ell N$, $\tau \to N N$ 
and $\tau \to \pi N$ channels.

\begin{figure}[h!]
	\centering
	\includegraphics[width=.45\textwidth]{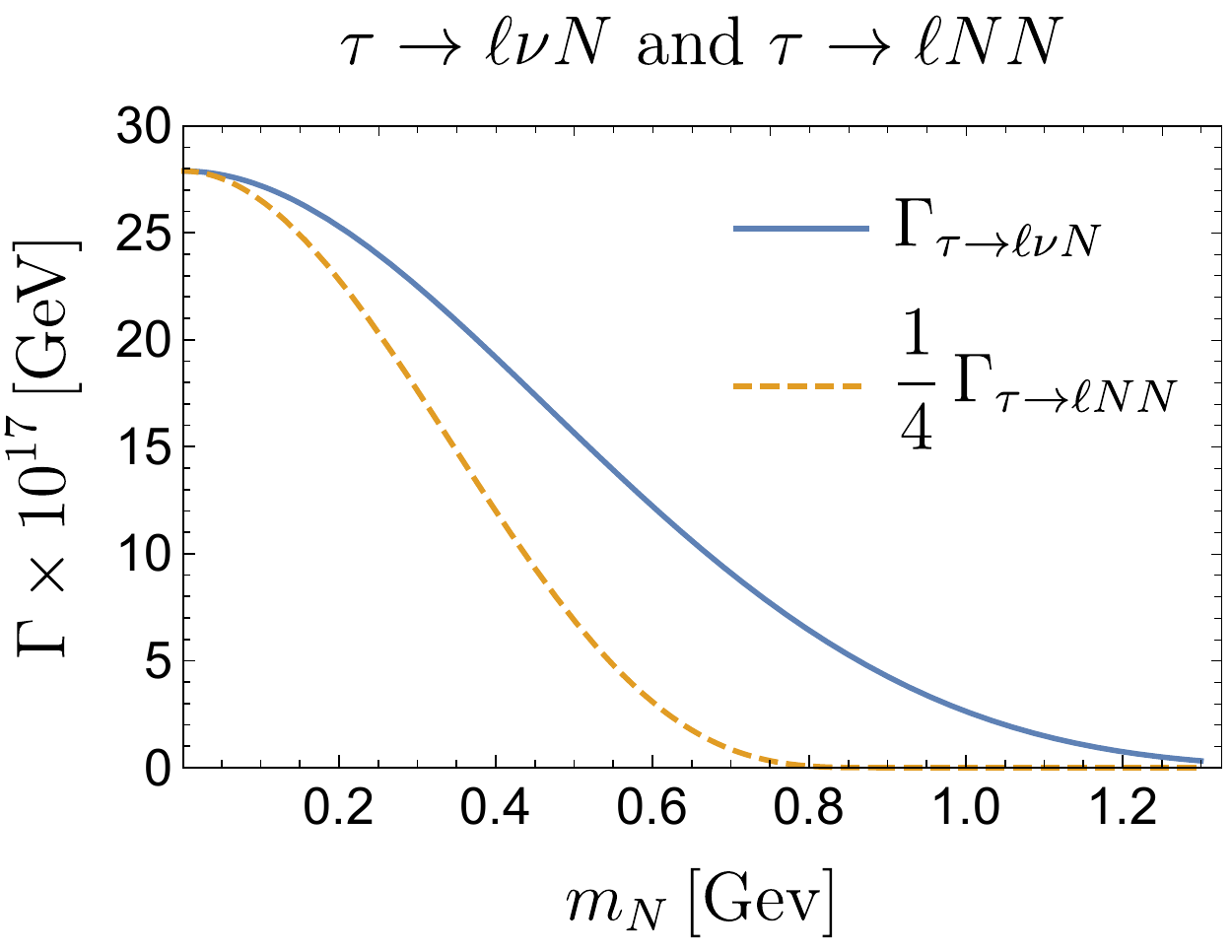}
	\quad
	\includegraphics[width=.45\textwidth]{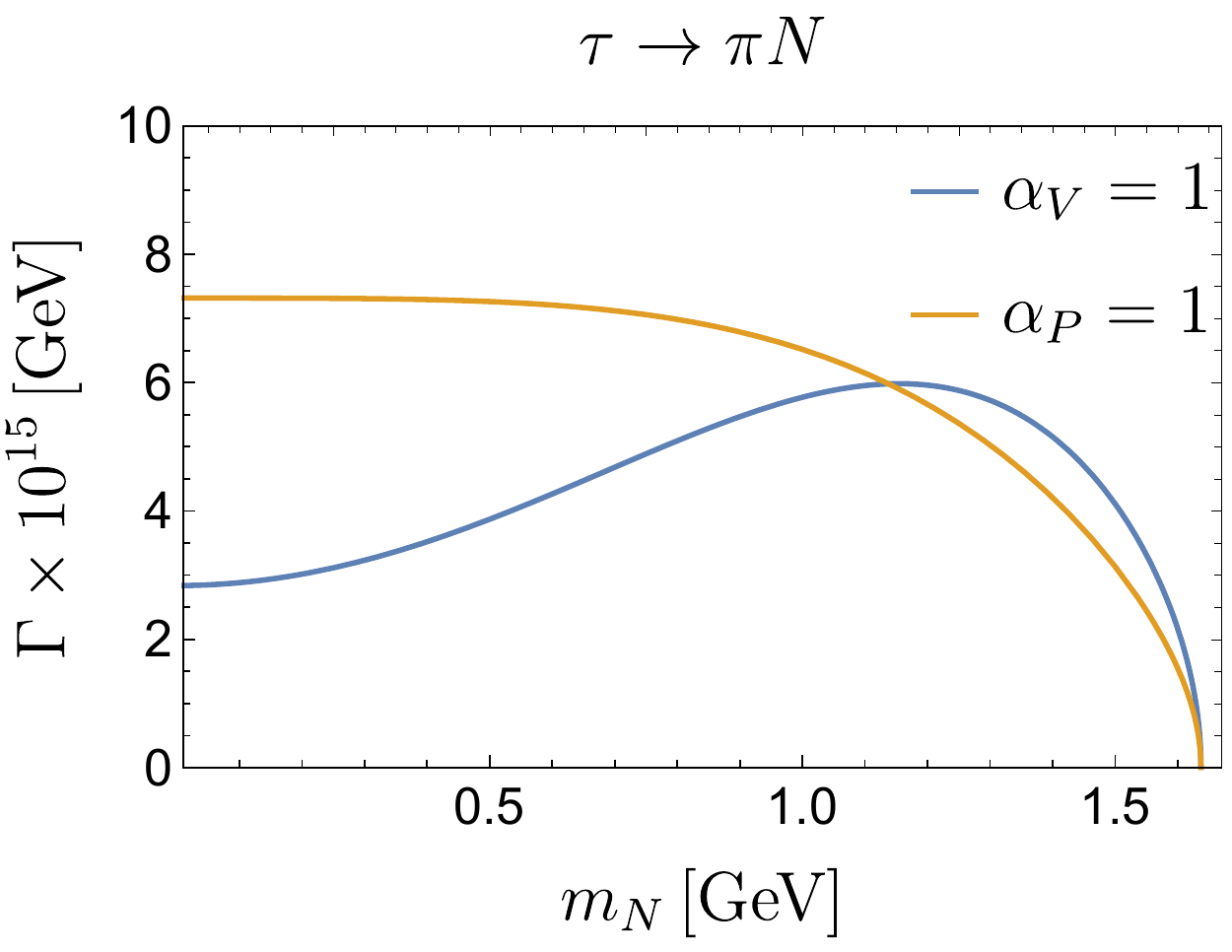}
	\caption{RH neutrino mass dependence of the $\tau$~decay width in different 
	decay channels. Left: Decay width of $\tau \to \ell \nu N$ and $\tau \to \ell N N$
	(rescaled) where
	the mass on the charged leptons has been neglected.
	Right: Decay width of $\tau \to \pi N$ for operators with axial and pseudo-scalar structures. }
	\label{fig:tauDecay_mN}
\end{figure}

\FloatBarrier
%\clearpage
\bibliographystyle{style}
\bibliography{draft}{}

\providecommand{\href}[2]{#2}\begingroup\raggedright\begin{thebibliography}{10}

\bibitem{Fukuda:1998mi}
{\scshape Super-Kamiokande} collaboration, Y.~Fukuda et~al., \emph{{Evidence
  for oscillation of atmospheric neutrinos}},
  \href{http://dx.doi.org/10.1103/PhysRevLett.81.1562}{\emph{Phys. Rev. Lett.}
  {\bf 81} (1998) 1562--1567},
  [\href{https://arxiv.org/abs/hep-ex/9807003}{{\tt hep-ex/9807003}}].

\bibitem{Toshito:2001dk}
{\scshape Super-Kamiokande} collaboration, T.~Toshito, \emph{{Super-Kamiokande
  atmospheric neutrino results}},  in \emph{{36th Rencontres de Moriond on
  Electroweak Interactions and Unified Theories}}, 5, 2001.
\newblock \href{https://arxiv.org/abs/hep-ex/0105023}{{\tt hep-ex/0105023}}.

\bibitem{Giacomelli:2001td}
{\scshape MACRO} collaboration, G.~Giacomelli and M.~Giorgini,
  \emph{{Atmospheric neutrino oscillations in MACRO}},  in \emph{{NO-VE
  International Workshop on Neutrino Oscillations in Venice}}, pp.~207--220,
  10, 2001.
\newblock \href{https://arxiv.org/abs/hep-ex/0110021}{{\tt hep-ex/0110021}}.

\bibitem{Fukuda:2001nj}
{\scshape Super-Kamiokande} collaboration, S.~Fukuda et~al., \emph{{Solar B-8
  and hep neutrino measurements from 1258 days of Super-Kamiokande data}},
  \href{http://dx.doi.org/10.1103/PhysRevLett.86.5651}{\emph{Phys. Rev. Lett.}
  {\bf 86} (2001) 5651--5655},
  [\href{https://arxiv.org/abs/hep-ex/0103032}{{\tt hep-ex/0103032}}].

\bibitem{Fukuda:2001nk}
{\scshape Super-Kamiokande} collaboration, S.~Fukuda et~al., \emph{{Constraints
  on neutrino oscillations using 1258 days of Super-Kamiokande solar neutrino
  data}}, \href{http://dx.doi.org/10.1103/PhysRevLett.86.5656}{\emph{Phys. Rev.
  Lett.} {\bf 86} (2001) 5656--5660},
  [\href{https://arxiv.org/abs/hep-ex/0103033}{{\tt hep-ex/0103033}}].

\bibitem{Ahmad:2002jz}
{\scshape SNO} collaboration, Q.~Ahmad et~al., \emph{{Direct evidence for
  neutrino flavor transformation from neutral current interactions in the
  Sudbury Neutrino Observatory}},
  \href{http://dx.doi.org/10.1103/PhysRevLett.89.011301}{\emph{Phys. Rev.
  Lett.} {\bf 89} (2002) 011301},
  [\href{https://arxiv.org/abs/nucl-ex/0204008}{{\tt nucl-ex/0204008}}].

\bibitem{Ahmad:2002ka}
{\scshape SNO} collaboration, Q.~Ahmad et~al., \emph{{Measurement of day and
  night neutrino energy spectra at SNO and constraints on neutrino mixing
  parameters}},
  \href{http://dx.doi.org/10.1103/PhysRevLett.89.011302}{\emph{Phys. Rev.
  Lett.} {\bf 89} (2002) 011302},
  [\href{https://arxiv.org/abs/nucl-ex/0204009}{{\tt nucl-ex/0204009}}].

\bibitem{Minkowski:1977sc}
P.~Minkowski, \emph{$\mu \rightarrow e \gamma$ at a rate of one out of
  $1$-billion muon decays?},
  \href{http://dx.doi.org/10.1016/0370-2693(77)90435-X}{\emph{Phys.Lett.} {\bf
  B67} (1977) 421}.

\bibitem{Yanagida:1979as}
T.~Yanagida, \emph{Horizontal symmetry and masses of neutrinos},
  {\emph{Conf.Proc.} {\bf C7902131} (1979) 95--99}.

\bibitem{GellMann:1980vs}
M.~Gell-Mann, P.~Ramond and R.~Slansky, \emph{Complex spinors and unified
  theories}, {\emph{Conf.Proc.} {\bf C790927} (1979) 315--321},
  [\href{https://arxiv.org/abs/1306.4669}{{\tt 1306.4669}}].

\bibitem{Mohapatra:1979ia}
R.~N. Mohapatra and G.~Senjanovic, \emph{Neutrino mass and spontaneous parity
  violation},
  \href{http://dx.doi.org/10.1103/PhysRevLett.44.912}{\emph{Phys.Rev.Lett.}
  {\bf 44} (1980) 912}.

\bibitem{tHooft:1979rat}
G.~'t~Hooft, \emph{{Naturalness, chiral symmetry, and spontaneous chiral
  symmetry breaking}},
  \href{http://dx.doi.org/10.1007/978-1-4684-7571-5\_9}{\emph{NATO Sci. Ser. B}
  {\bf 59} (1980) 135--157}.

\bibitem{Pilaftsis:1991ug}
A.~Pilaftsis, \emph{{Radiatively induced neutrino masses and large Higgs
  neutrino couplings in the standard model with Majorana fields}},
  \href{http://dx.doi.org/10.1007/BF01482590}{\emph{Z. Phys. C} {\bf 55} (1992)
  275--282}, [\href{https://arxiv.org/abs/hep-ph/9901206}{{\tt
  hep-ph/9901206}}].

\bibitem{Borzumati:2000mc}
F.~Borzumati and Y.~Nomura, \emph{{Low scale seesaw mechanisms for light
  neutrinos}}, \href{http://dx.doi.org/10.1103/PhysRevD.64.053005}{\emph{Phys.
  Rev. D} {\bf 64} (2001) 053005},
  [\href{https://arxiv.org/abs/hep-ph/0007018}{{\tt hep-ph/0007018}}].

\bibitem{Dev:2016dja}
P.~S.~B. Dev, R.~N. Mohapatra and Y.~Zhang, \emph{{Probing the Higgs Sector of
  the Minimal Left-Right Symmetric Model at Future Hadron Colliders}},
  \href{http://dx.doi.org/10.1007/JHEP05(2016)174}{\emph{JHEP} {\bf 05} (2016)
  174}, [\href{https://arxiv.org/abs/1602.05947}{{\tt 1602.05947}}].

\bibitem{Dev:2009aw}
P.~Dev and R.~Mohapatra, \emph{{TeV Scale Inverse Seesaw in SO(10) and Leptonic
  Non-Unitarity Effects}},
  \href{http://dx.doi.org/10.1103/PhysRevD.81.013001}{\emph{Phys. Rev. D} {\bf
  81} (2010) 013001}, [\href{https://arxiv.org/abs/0910.3924}{{\tt
  0910.3924}}].

\bibitem{Borzumati:1986qx}
F.~Borzumati and A.~Masiero, \emph{{Large Muon and electron Number Violations
  in Supergravity Theories}},
  \href{http://dx.doi.org/10.1103/PhysRevLett.57.961}{\emph{Phys. Rev. Lett.}
  {\bf 57} (1986) 961}.

\bibitem{delAguila:2008ir}
F.~del Aguila, S.~Bar-Shalom, A.~Soni and J.~Wudka, \emph{{Heavy Majorana
  Neutrinos in the Effective Lagrangian Description: Application to Hadron
  Colliders}},
  \href{http://dx.doi.org/10.1016/j.physletb.2008.11.031}{\emph{Phys. Lett.}
  {\bf B670} (2009) 399--402}, [\href{https://arxiv.org/abs/0806.0876}{{\tt
  0806.0876}}].

\bibitem{Aparici:2009fh}
A.~Aparici, K.~Kim, A.~Santamaria and J.~Wudka, \emph{{Right-handed neutrino
  magnetic moments}},
  \href{http://dx.doi.org/10.1103/PhysRevD.80.013010}{\emph{Phys. Rev.} {\bf
  D80} (2009) 013010}, [\href{https://arxiv.org/abs/0904.3244}{{\tt
  0904.3244}}].

\bibitem{Bhattacharya:2015vja}
S.~Bhattacharya and J.~Wudka, \emph{{Dimension-seven operators in the standard
  model with right handed neutrinos}},
  \href{http://dx.doi.org/10.1103/PhysRevD.94.055022,
  10.1103/PhysRevD.95.039904}{\emph{Phys. Rev.} {\bf D94} (2016) 055022},
  [\href{https://arxiv.org/abs/1505.05264}{{\tt 1505.05264}}].

\bibitem{Liao:2016qyd}
Y.~Liao and X.-D. Ma, \emph{{Operators up to Dimension Seven in Standard Model
  Effective Field Theory Extended with Sterile Neutrinos}},
  \href{http://dx.doi.org/10.1103/PhysRevD.96.015012}{\emph{Phys. Rev.} {\bf
  D96} (2017) 015012}, [\href{https://arxiv.org/abs/1612.04527}{{\tt
  1612.04527}}].

\bibitem{Chala:2020vqp}
M.~Chala and A.~Titov, \emph{{One-loop matching in the SMEFT extended with a
  sterile neutrino}},  \href{https://arxiv.org/abs/2001.07732}{{\tt
  2001.07732}}.

\bibitem{Dekens:2020ttz}
W.~Dekens, J.~de~Vries, K.~Fuyuto, E.~Mereghetti and G.~Zhou, \emph{{Sterile
  neutrinos and neutrinoless double beta decay in effective field theory}},
  \href{https://arxiv.org/abs/2002.07182}{{\tt 2002.07182}}.

\bibitem{Cai:2017mow}
Y.~Cai, T.~Han, T.~Li and R.~Ruiz, \emph{{Lepton Number Violation: Seesaw
  Models and Their Collider Tests}},
  \href{http://dx.doi.org/10.3389/fphy.2018.00040}{\emph{Front. in Phys.} {\bf
  6} (2018) 40}, [\href{https://arxiv.org/abs/1711.02180}{{\tt 1711.02180}}].

\bibitem{Duarte:2015iba}
L.~Duarte, J.~Peressutti and O.~A. Sampayo, \emph{{Majorana neutrino decay in
  an Effective Approach}},
  \href{http://dx.doi.org/10.1103/PhysRevD.92.093002}{\emph{Phys. Rev.} {\bf
  D92} (2015) 093002}, [\href{https://arxiv.org/abs/1508.01588}{{\tt
  1508.01588}}].

\bibitem{Butterworth:2019iff}
J.~M. Butterworth, M.~Chala, C.~Englert, M.~Spannowsky and A.~Titov,
  \emph{{Higgs phenomenology as a probe of sterile neutrinos}},
  \href{http://dx.doi.org/10.1103/PhysRevD.100.115019}{\emph{Phys. Rev.} {\bf
  D100} (2019) 115019}, [\href{https://arxiv.org/abs/1909.04665}{{\tt
  1909.04665}}].

\bibitem{Duarte:2016caz}
L.~Duarte, J.~Peressutti and O.~A. Sampayo, \emph{{Not-that-heavy Majorana
  neutrino signals at the LHC}},
  \href{http://dx.doi.org/10.1088/1361-6471/aa99f5}{\emph{J. Phys.} {\bf G45}
  (2018) 025001}, [\href{https://arxiv.org/abs/1610.03894}{{\tt 1610.03894}}].

\bibitem{Cirigliano:2005ck}
V.~Cirigliano, B.~Grinstein, G.~Isidori and M.~B. Wise, \emph{{Minimal flavor
  violation in the lepton sector}},
  \href{http://dx.doi.org/10.1016/j.nuclphysb.2005.08.037}{\emph{Nucl. Phys. B}
  {\bf 728} (2005) 121--134}, [\href{https://arxiv.org/abs/hep-ph/0507001}{{\tt
  hep-ph/0507001}}].

\bibitem{Canas:2015yoa}
B.~C. Canas, O.~G. Miranda, A.~Parada, M.~Tortola and J.~W.~F. Valle,
  \emph{{Updating neutrino magnetic moment constraints}},
  \href{http://dx.doi.org/10.1016/j.physletb.2016.03.078,
  10.1016/j.physletb.2015.12.011}{\emph{Phys. Lett.} {\bf B753} (2016)
  191--198}, [\href{https://arxiv.org/abs/1510.01684}{{\tt 1510.01684}}].

\bibitem{Miranda:2019wdy}
O.~G. Miranda, D.~K. Papoulias, M.~T{\'o}rtola and J.~W.~F. Valle,
  \emph{{Probing neutrino transition magnetic moments with coherent elastic
  neutrino-nucleus scattering}},  \href{https://arxiv.org/abs/1905.03750}{{\tt
  1905.03750}}.

\bibitem{Casas:2001sr}
J.~Casas and A.~Ibarra, \emph{{Oscillating neutrinos and $\mu \to e, \gamma$}},
  \href{http://dx.doi.org/10.1016/S0550-3213(01)00475-8}{\emph{Nucl. Phys. B}
  {\bf 618} (2001) 171--204}, [\href{https://arxiv.org/abs/hep-ph/0103065}{{\tt
  hep-ph/0103065}}].

\bibitem{Barducci:2020ncz}
D.~Barducci, E.~Bertuzzo, A.~Caputo and P.~Hernandez, \emph{{Minimal flavor
  violation in the see-saw portal}},
  \href{https://arxiv.org/abs/2003.08391}{{\tt 2003.08391}}.

\bibitem{Acciarri:1997im}
{\scshape L3} collaboration, M.~Acciarri et~al., \emph{{Search for new physics
  in energetic single photon production in $e^{+} e^{-}$ annihilation at the
  $Z$ resonance}},
  \href{http://dx.doi.org/10.1016/S0370-2693(97)01003-4}{\emph{Phys. Lett.}
  {\bf B412} (1997) 201--209}.

\bibitem{Tanabashi:2018oca}
{\scshape Particle Data Group} collaboration, M.~Tanabashi et~al.,
  \emph{{Review of Particle Physics}},
  \href{http://dx.doi.org/10.1103/PhysRevD.98.030001}{\emph{Phys. Rev.} {\bf
  D98} (2018) 030001}.

\bibitem{Chala:2020pbn}
M.~Chala and A.~Titov, \emph{{One-loop running of dimension-six Higgs-neutrino
  operators and implications of a large neutrino dipole moment}},
  \href{https://arxiv.org/abs/2006.14596}{{\tt 2006.14596}}.

\bibitem{Alwall:2014hca}
J.~Alwall, R.~Frederix, S.~Frixione, V.~Hirschi, F.~Maltoni, O.~Mattelaer
  et~al., \emph{{The automated computation of tree-level and next-to-leading
  order differential cross sections, and their matching to parton shower
  simulations}}, \href{http://dx.doi.org/10.1007/JHEP07(2014)079}{\emph{JHEP}
  {\bf 07} (2014) 079}, [\href{https://arxiv.org/abs/1405.0301}{{\tt
  1405.0301}}].

\bibitem{Buckley:2014ana}
A.~Buckley, J.~Ferrando, S.~Lloyd, K.~Nordström, B.~Page, M.~Rüfenacht
  et~al., \emph{{LHAPDF6: parton density access in the LHC precision era}},
  \href{http://dx.doi.org/10.1140/epjc/s10052-015-3318-8}{\emph{Eur. Phys. J.
  C} {\bf 75} (2015) 132}, [\href{https://arxiv.org/abs/1412.7420}{{\tt
  1412.7420}}].

\bibitem{Sjostrand:2014zea}
T.~Sj{\"o}strand, S.~Ask, J.~R. Christiansen, R.~Corke, N.~Desai, P.~Ilten
  et~al., \emph{{An Introduction to PYTHIA 8.2}},
  \href{http://dx.doi.org/10.1016/j.cpc.2015.01.024}{\emph{Comput. Phys.
  Commun.} {\bf 191} (2015) 159--177},
  [\href{https://arxiv.org/abs/1410.3012}{{\tt 1410.3012}}].

\bibitem{Dobbs:2001ck}
M.~Dobbs and J.~B. Hansen, \emph{{The HepMC C++ Monte Carlo event record for
  High Energy Physics}},
  \href{http://dx.doi.org/10.1016/S0010-4655(00)00189-2}{\emph{Comput. Phys.
  Commun.} {\bf 134} (2001) 41--46}.

\bibitem{Cacciari:2011ma}
M.~Cacciari, G.~P. Salam and G.~Soyez, \emph{{FastJet User Manual}},
  \href{http://dx.doi.org/10.1140/epjc/s10052-012-1896-2}{\emph{Eur. Phys. J.}
  {\bf C72} (2012) 1896}, [\href{https://arxiv.org/abs/1111.6097}{{\tt
  1111.6097}}].

\bibitem{Alcaide:2019pnf}
J.~Alcaide, S.~Banerjee, M.~Chala and A.~Titov, \emph{{Probes of the Standard
  Model effective field theory extended with a right-handed neutrino}},
  \href{http://dx.doi.org/10.1007/JHEP08(2019)031}{\emph{JHEP} {\bf 08} (2019)
  031}, [\href{https://arxiv.org/abs/1905.11375}{{\tt 1905.11375}}].

\bibitem{Sirunyan:2018psa}
{\scshape CMS} collaboration, A.~M. Sirunyan et~al., \emph{{Search for
  supersymmetry in events with a photon, a lepton, and missing transverse
  momentum in proton-proton collisions at $\sqrt{s} =$ 13 TeV}},
  \href{http://dx.doi.org/10.1007/JHEP01(2019)154}{\emph{JHEP} {\bf 01} (2019)
  154}, [\href{https://arxiv.org/abs/1812.04066}{{\tt 1812.04066}}].

\bibitem{Aad:2014fha}
{\scshape ATLAS} collaboration, G.~Aad et~al., \emph{{Search for new resonances
  in $W\gamma$ and $Z\gamma$ final states in $pp$ collisions at $\sqrt s=8$ TeV
  with the ATLAS detector}},
  \href{http://dx.doi.org/10.1016/j.physletb.2014.10.002}{\emph{Phys.\ Lett.\
  B} {\bf 738} (2014) 428--447}, [\href{https://arxiv.org/abs/1407.8150}{{\tt
  1407.8150}}].

\bibitem{Read:2002hq}
A.~L. Read, \emph{{Presentation of search results: The CL(s) technique}},
  \href{http://dx.doi.org/10.1088/0954-3899/28/10/313}{\emph{J. Phys.} {\bf
  G28} (2002) 2693--2704}.

\bibitem{Sirunyan:2019mbp}
{\scshape CMS} collaboration, A.~M. Sirunyan et~al., \emph{{Search for
  supersymmetry in final states with photons and missing transverse momentum in
  proton-proton collisions at 13 TeV}},
  \href{http://dx.doi.org/10.1007/JHEP06(2019)143}{\emph{JHEP} {\bf 06} (2019)
  143}, [\href{https://arxiv.org/abs/1903.07070}{{\tt 1903.07070}}].

\bibitem{Carpentier:2010ue}
M.~Carpentier and S.~Davidson, \emph{{Constraints on two-lepton, two quark
  operators}},
  \href{http://dx.doi.org/10.1140/epjc/s10052-010-1482-4}{\emph{Eur. Phys. J.}
  {\bf C70} (2010) 1071--1090}, [\href{https://arxiv.org/abs/1008.0280}{{\tt
  1008.0280}}].

\bibitem{Bychkov:2008ws}
M.~Bychkov et~al., \emph{{New Precise Measurement of the Pion Weak Form Factors
  in pi+ ---> e+ nu gamma Decay}},
  \href{http://dx.doi.org/10.1103/PhysRevLett.103.051802}{\emph{Phys. Rev.
  Lett.} {\bf 103} (2009) 051802}, [\href{https://arxiv.org/abs/0804.1815}{{\tt
  0804.1815}}].

\bibitem{Bressi:1997gs}
G.~Bressi, G.~Carugno, E.~Conti, A.~Meneguzzo, S.~Cerdonio and D.~Zanello,
  \emph{{New measurement of the pi --> mu nu gamma decay}},
  \href{http://dx.doi.org/10.1016/S0550-3213(97)00734-7}{\emph{Nucl.\ Phys.\ B}
  {\bf 513} (1998) 555--572}.

\bibitem{Fael:2015gua}
M.~Fael, L.~Mercolli and M.~Passera, \emph{{Radiative $\mu$ and $\tau$ leptonic
  decays at NLO}}, \href{http://dx.doi.org/10.1007/JHEP07(2015)153}{\emph{JHEP}
  {\bf 07} (2015) 153}, [\href{https://arxiv.org/abs/1506.03416}{{\tt
  1506.03416}}].

\bibitem{Achard:2003tx}
{\scshape L3} collaboration, P.~Achard et~al., \emph{{Single photon and
  multiphoton events with missing energy in $e^{+} e^{-}$ collisions at LEP}},
  \href{http://dx.doi.org/10.1016/j.physletb.2004.01.010}{\emph{Phys. Lett. B}
  {\bf 587} (2004) 16--32}, [\href{https://arxiv.org/abs/hep-ex/0402002}{{\tt
  hep-ex/0402002}}].

\bibitem{Bischer:2019ttk}
I.~Bischer and W.~Rodejohann, \emph{{General neutrino interactions from an
  effective field theory perspective}},
  \href{http://dx.doi.org/10.1016/j.nuclphysb.2019.114746}{\emph{Nucl. Phys. B}
  {\bf 947} (2019) 114746}, [\href{https://arxiv.org/abs/1905.08699}{{\tt
  1905.08699}}].

\bibitem{Miyazaki:2007jp}
{\scshape Belle} collaboration, Y.~Miyazaki et~al., \emph{{Search for lepton
  flavor violating tau- decays into l- eta, l- eta-prime and l- pi0}},
  \href{http://dx.doi.org/10.1016/j.physletb.2007.03.027}{\emph{Phys. Lett. B}
  {\bf 648} (2007) 341--350}, [\href{https://arxiv.org/abs/hep-ex/0703009}{{\tt
  hep-ex/0703009}}].

\bibitem{Cacciari:2008gp}
M.~Cacciari, G.~P. Salam and G.~Soyez, \emph{{The anti-$k_t$ jet clustering
  algorithm}},
  \href{http://dx.doi.org/10.1088/1126-6708/2008/04/063}{\emph{JHEP} {\bf 04}
  (2008) 063}, [\href{https://arxiv.org/abs/0802.1189}{{\tt 0802.1189}}].

\bibitem{Liu:2019qfa}
N.~Liu, Z.-G. Si, L.~Wu, H.~Zhou and B.~Zhu, \emph{{Top quark as a probe of
  heavy Majorana neutrino at the LHC and future colliders}},
  \href{http://dx.doi.org/10.1103/PhysRevD.101.071701}{\emph{Phys. Rev. D} {\bf
  101} (2020) 071701}, [\href{https://arxiv.org/abs/1910.00749}{{\tt
  1910.00749}}].

\end{thebibliography}\endgroup

\end{document}